\documentclass[aoas]{imsart}

\RequirePackage[OT1]{fontenc}
\RequirePackage{amsthm,amsmath,amsfonts,amssymb}
\RequirePackage[authoryear]{natbib}
\RequirePackage[colorlinks,citecolor=blue,urlcolor=blue]{hyperref}

\startlocaldefs
\numberwithin{equation}{section}
\theoremstyle{plain}

\endlocaldefs

\makeatletter

\usepackage[outerbars,color]{changebar}
\ifx\pdfoutput\undefined
\else\ifnum\pdfoutput>0
  \usepackage{pdfcolmk}
\fi\fi
\cbcolor{black}

\usepackage{listings}
\usepackage{mathtools}

\newcommand{\mbP}{\mathbb{P}}
\newcommand{\s}{\vspace{0.25cm}}
\newcommand{\bZ}{\bm{Z}}
\newcommand{\mR}{\mathbb{R}}
\usepackage{tikz}
\usetikzlibrary{bayesnet}

\usetikzlibrary{fit,positioning}
\newcommand{\obs}{latent}
\newcommand{\lat}{obs}
\tikzstyle{latentf} = [rectangle,fill=white,draw=black,inner sep=1pt,
\tikzstyle{obf} = [latent,fill=gray!25]

\usepackage{bm}

\newcommand{\beno}{\begin{equation}\begin{array}{lllllllllllll}\nonumber}
\newcommand{\ee}{\end{array}\end{equation}}

\usepackage{changepage}
\usepackage[normalem]{ulem}
\usepackage[final]{pdfpages}
\usepackage{rotating}
\usepackage{lscape}
\usepackage{pdflscape}
\usepackage{afterpage}
\usepackage{longtable}
\usepackage{caption}
\usepackage{subcaption}
\usepackage{array}
\newcolumntype{L}{>{\centering\arraybackslash}m{3cm}}
\newcommand{\msim}{\mathop{\rm \sim}}
\newcommand{\ind}{\msim\limits^{\mbox{\tiny ind}}}
\newcommand{\iid}{\msim\limits^{\mbox{\tiny iid}}}

\usepackage{xcolor}
\colorlet{linkequation}{blue}

\usepackage{graphicx}

\usepackage{multicol}
\usepackage{multirow}

\usepackage[english]{babel}

\newcommand{\hide}[1]{}
\newcommand{\logit}{\mbox{logit}}
\newcommand{\bz}{\bm{z}}

\usepackage[scr=rsfs]{mathalpha}
\newcommand{\mD}{\mathscr{D}}
\begin{document}

\begin{frontmatter}

\title{A Socio-Demographic Latent Space Approach to Spatial Data When Geography is Important but not All-Important}

\runtitle{Latent Socio-Spatial Space Model}

\begin{aug}
\runauthor{Nandy et al.}
\author[A]{\fnms{Saikat}~\snm{Nandy}\ead[label=e1]{snandy@missouri.edu}\orcid{0000-0003-4725-8773}},
\author[A,B]{\fnms{Scott~H.}~\snm{Holan}\ead[label=e2,mark]{holans@missouri.edu}\ead[label=e3,mark]{scott.holan@census.gov}},
\and
\author[C]{\fnms{Michael}~\snm{Schweinberger}\ead[label=e4,mark]{mus47@psu.edu}}

\address[A]{Department of Statistics, University of Missouri\printead[presep={,\ }]{e1,e2}}
\address[B]{Office of the Associate Director for Research and Methodology, U.S. Census Bureau\printead[presep={,\ }]{e3}}
\address[C]{Department of Statistics, The Pennsylvania State University\printead[presep={,\ }]{e4}}
\end{aug}

\begin{abstract}
Many models for spatial and spatio-temporal data assume that ``near things are more related
than distant things," which is known as the first law of geography. While geography may be important, it may not be all-important, for at least two reasons. First, technology helps bridge distance, so that regions separated by large distances may be more similar than would be expected based on geographical distance. Second, geographical, political, and social divisions can make neighboring regions dissimilar. We develop a flexible Bayesian approach for learning from spatial data which units are close in an unobserved socio-demographic space and hence which units are similar. As a by-product, the Bayesian approach helps quantify the relative importance of socio-demographic space relative to geographical space. To demonstrate the proposed approach, we present simulations along with an application to county-level data on median household income in the U.S.\ state of Florida.
\end{abstract}



\end{frontmatter}


\section{Introduction}
\label{sec:introduction}

All observations are spatial and temporal,
in the sense that all observations are made at specific points in space and time \citep{ChBaKuMaPe22}.
In scenarios in which space and time are believed to affect real-world phenomena of interest,
it is often assumed that {\em ``near things are more related than distant things,"}
which is known as the first law of geography \citep{Tobler1970}.

While geography may be important,
it may not be all-important,
for at least two reasons.
First,
technology helps bridge distance,
so that events in one region can be affected by events in distant regions.
As a consequence,
regions separated by large distances may be more similar than would be expected based on geographical distance.
Second,
geographical, political, and social divisions can make neighboring regions dissimilar.
For example,
mountains and other geographical features can divide communities,
giving rise to communities that are more dissimilar than would be expected based on geographical distance:
e.g., 
the shortest route between the two mountain towns of Aspen and Crested Butte in the U.S.\ state of Colorado is less than 40 miles long,
but the fact that these towns are separated by a wall of 13,000--14,000 feet high mountains has led to the development of two towns that are dissimilar in terms of cost of living and other socio-demographic characteristics.
In addition to geographical features,
borders between states can divide communities that are close in terms of geographical distance and can therefore make them dissimilar rather than similar.
Last, 
but not least,
racial segregation and social divisions can give rise to neighborhoods that are close in terms of geographical distance but are dissimilar in terms of race, 
access to medical care, 
educational opportunities, and income \citep{Chetty2018}. 
We focus here on the second problem rather than the first,
and defer solutions to the first problem to future research.

\subsection{Motivating example}

As a case in point, 
we consider areal data collected by the American Community Survey (ACS) of the U.S.\ Census Bureau,
released in 2019.
The 2019 ACS data contains public-use five-year period design-based estimates of demographic variables at the county level in the U.S.\ states of Florida, Georgia, 
North and South Carolina,
and other U.S.\ states.
\begin{figure}[t]
\begin{subfigure}{\columnwidth}
    \includegraphics[width=0.5\columnwidth]{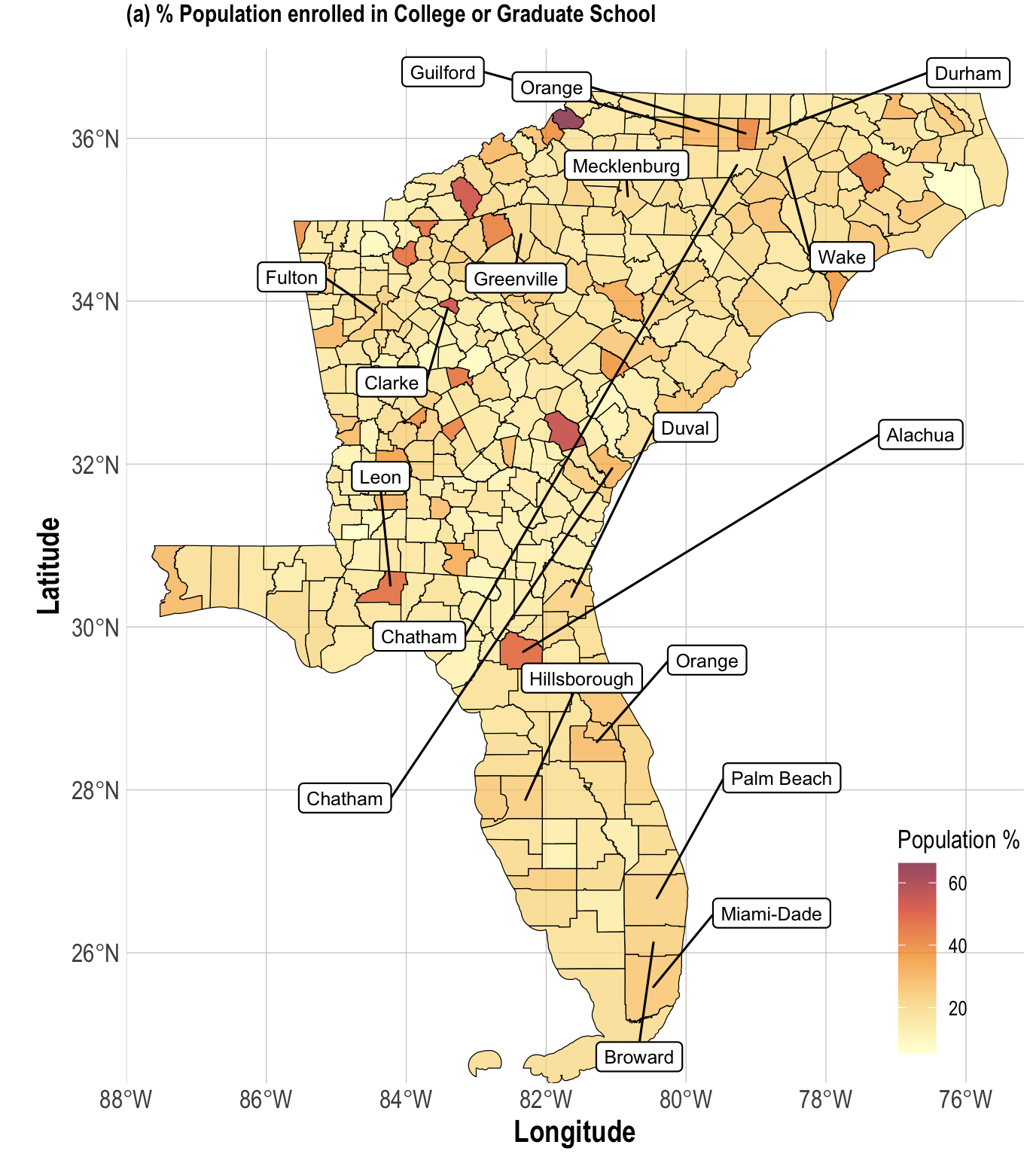}
    \includegraphics[width=0.5\columnwidth]{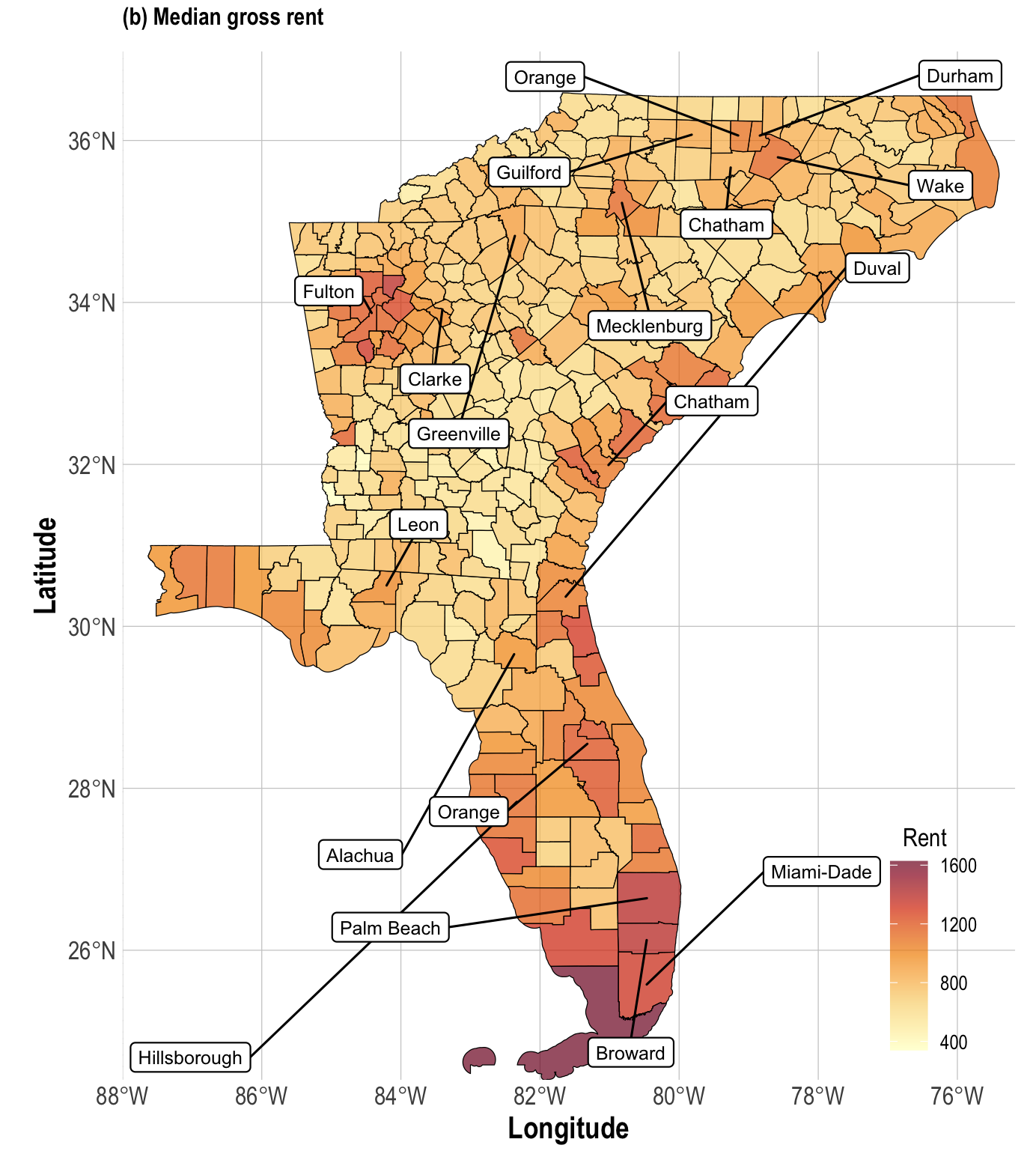}
\hfill{}
\end{subfigure} \\
\begin{subfigure}{\linewidth}
    \includegraphics[width=0.5\linewidth]{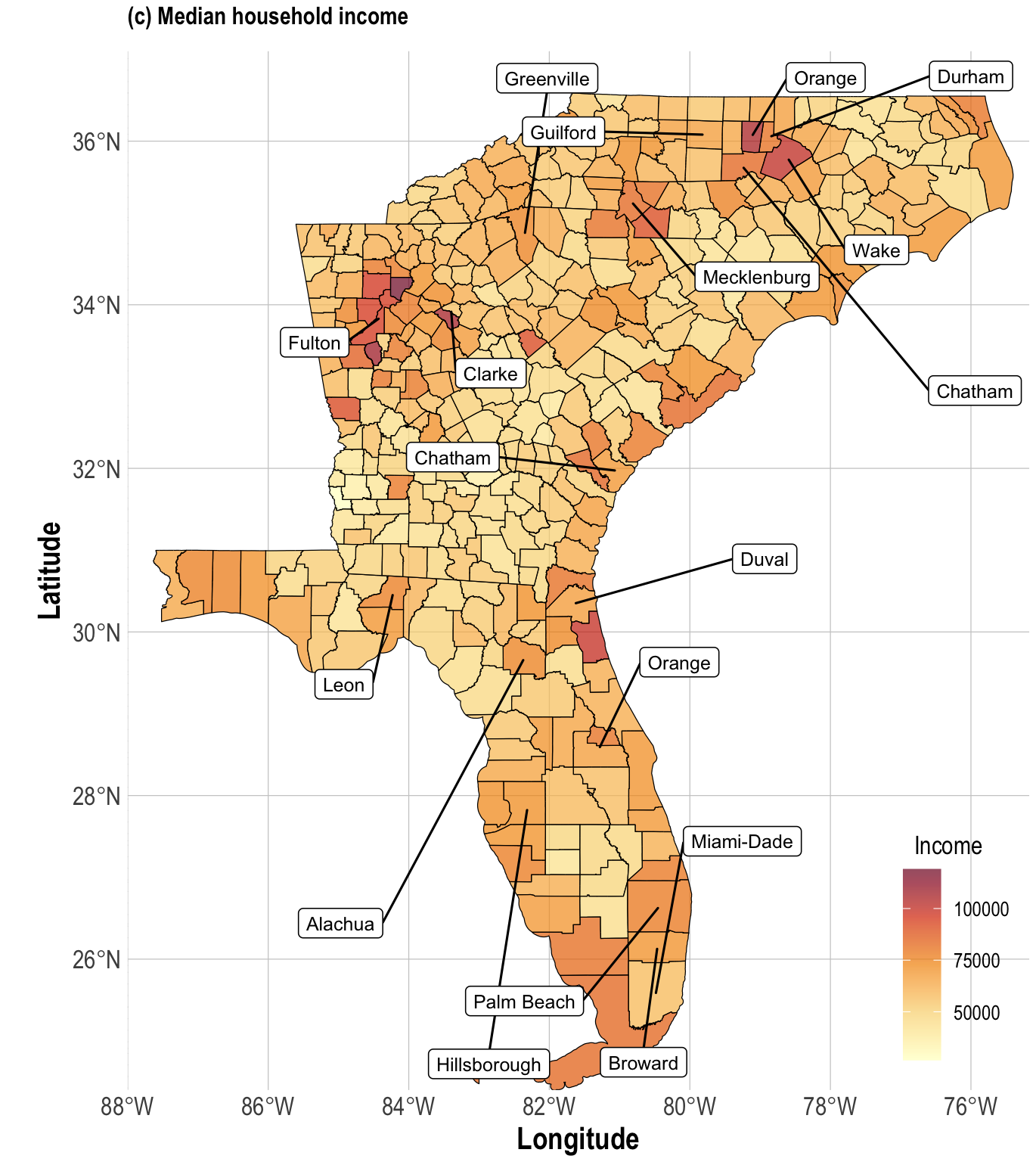}
    \includegraphics[width=0.5\linewidth]{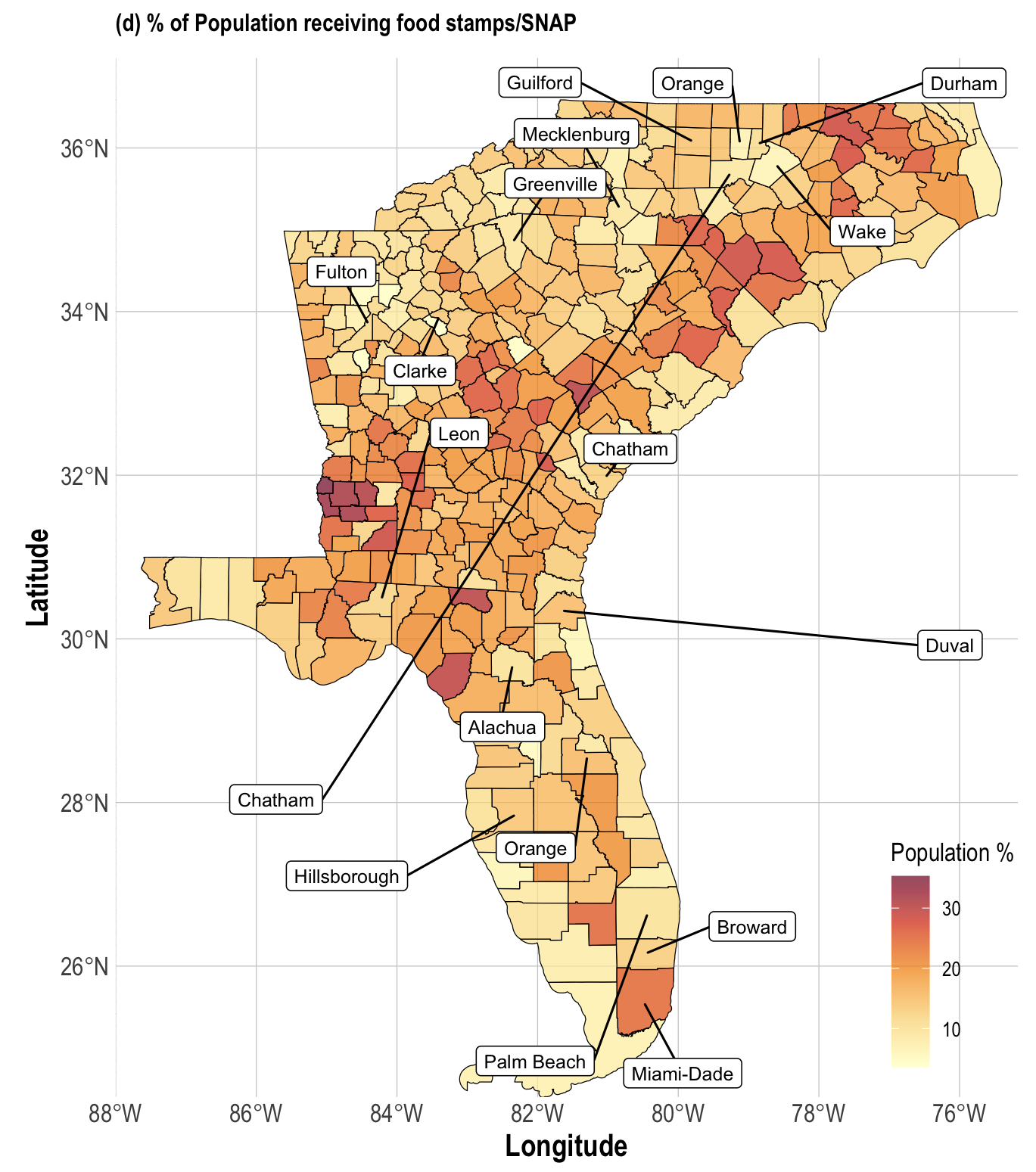}
\end{subfigure}
\caption{2019 ACS design-based estimates from counties in Florida, Georgia, 
North and South Carolina: 
(a) the percentage of population enrolled in college or graduate school;
(b) median gross rent; 
(c) median household income;
and
(d) the percentage of population receiving food assistance in the form of food stamps or SNAP.}
    \label{fig: introData}

\end{figure}
Out of the 372 counties in the four mentioned U.S.\ states,
we highlight sixteen counties that stand out: 
Miami-Dade, Broward, Palm Beach, Orange, Hillsborough, Leon, Alachual, and Duval counties in Florida;
Fulton, Clarke, and Clatham counties in Georgia;
Greenville county in South Carolina; 
and Durham, Mecklenburg, Guilford, and Wake counties in North Carolina.
These counties include some of the most populous metropolitan areas in the selected geographical area:
Miami, Fort Lauderdale, Orlando, Tampa, and Jacksonville in Florida;
Atlanta, Athens, and Savannah in Georgia;
and Raleigh, Chapel Hill, and Greensboro in North Carolina. 
Among them are so-called ``college towns,"
with a socio-demographic makeup dominated by a more diverse and younger population compared with other counties in the selected geographical area.
The first of the four maps in Figure~\ref{fig: introData} reveals that the counties with the highest percentage of college or graduate school enrollment are scattered throughout the geographical area of interest.
Some of the counties with the highest enrollment (e.g., Leon and Alachua counties in Florida, Clarke County in Georgia, and Orange County in North Carolina) are home to popular universities,
but are surrounded by counties with far lower levels of enrollment.
The second map in Figure~\ref{fig: introData} shows that counties along the coast report higher median gross rent than the counties in the interior, 
with the exception of metropolitan areas such as Atlanta in Georgia and the Research Triangle in North Carolina. 
The third and fourth map in Figure~\ref{fig: introData} underscore the fact that these counties are dissimilar from geographical neighbors when markers of economic conditions are considered, 
such as median household income or the percentage of population receiving food assistance in the form of food stamps or SNAP (Supplemental Nutrition Assistance Program).

\subsection{Existing and proposed approaches}
\label{sec:neighborhoods}

Many existing models for spatial and spatio-temporal data assume that {\em ``near things are more related than distant things,"}
in line with the first law of geography (e.g., see \citep{WiZMCr19} and the references therein).
For example,
many models encourage units that are geographic neighbors to be similar.
Nearest-neighbor approaches to spatial data were introduced by \cite{Whittle1954} and \cite{Bartlett1967} and form the basis of popular statistical models for capturing spatial dependencies, 
such as simultaneous autoregressive (SAR) models \citep{Whittle1954}, conditionally autoregressive (CAR) models \citep{Besag1974}, or intrinsic conditionally autoregressive (ICAR) models \citep{BesagYorkMollie1991}. 
These popular spatial models construct a first-order $N \times N$ neighborhood matrix $\mathbf{A}$, 
with $N$ being the number of units in the spatial domain.
The entries $A_{i,j} \in \{0, 1\}$ of the matrix $\mathbf{A}$ are binary, 
where $A_{i,j}=1$ if spatial units $i$ and $j$ share a geographic boundary and are therefore considered neighbors, 
and $A_{i,j}=0$ otherwise. 
Neighboring units are then encouraged to be similar.
\citet{Chaskin1997}, \citet{Grannis1998, Grannis2005}, \citet{Galster2001}, and \cite{Cutchin2011} introduced alternative definitions of neighborhoods,
motivated by social or ecological considerations.

Despite these and other well-established definitions of neighborhoods,
there is no universal definition of neighborhoods.
In the absence of a universal definition of neighborhoods,
it makes sense to learn the neighborhoods of units from spatial data.
In scenarios with independent replications from the same source being available, 
high-dimensional model selection methods for Markov random fields can be used to recover the neighborhoods of statistical units under strong assumptions \citep[e.g.,][]{MeBu06, Ravikumar2010}. 
That said,
recovering the neighborhoods of statistical units is non-trivial in scenarios without independent replications,
which arise in spatial and spatio-temporal statistics.
In scenarios without independent replications,
first steps towards estimating the neighborhoods of statistical units were made by \citet{JiSe96}, \citet{Csiszar2006}, \citet{White2009}, and \citet{GaO2019} with a focus on spatial data,
and \citet{Schweinberger2017} with a focus on spatio-temporal data.
For example,
\citet{JiSe96} and \citet{Csiszar2006} estimate the neighborhoods of discrete Markov random fields,
but those procedures are limited to discrete Markov random fields on a $d$-dimensional lattice $\mathbb{Z}^d$ and the neighborhoods of units are assumed to be symmetric sets. 
Bayesian areal wombling literature adopt a fully  Bayesian approach where they treat the edge elements of the adjacency matrix as random quantities to be updated along with other model parameters \citep{Lu2007,Ma2010}.
\citet{GaO2019} explored an alternative approach to estimating neighborhoods from spatial data by leveraging covariates,
but the quality of statistical conclusions may depend on the quality of the covariates at hand.

By contrast,
we introduce a novel statistical approach to learning from spatial data which units are close in an unobserved socio-demographic space.
As a by-product,
we quantify the relative importance of socio-demographic space relative to geographical space, 
along with the uncertainty associated with it.

\subsection{Structure of the paper}

We first review the 2019 ACS  data that motivated the proposed statistical framework in Section~\ref{section: Data} and then introduce the statistical framework in Section~\ref{section: Methodology}.
Simulation results and an application to the 2019 ACS  data are presented in Sections~\ref{section: Simulations} and~\ref{section: RealEg}, 
respectively. 
Throughout,
we adopt the statistical notations that are prevalent in the network statistics literature rather than the spatial statistics literature.
 
\section{American Community Survey Data} 
\label{section: Data}

To motivate the proposed statistical framework,
we leverage the ACS public-use data on demographic, economic, social and other quantities of interest at various levels of geographical granularity,
released by the U.S.\ Census Bureau in 2019.
The ACS provides design-based estimates of population quantities over a one-year period for geographical areas with at least 65,000 people and combines five consecutive years of ACS data to produce five-year period design-based estimates for geographical areas with at least 20,000 residents (\cite{Census2020}; \url{https://data.census.gov}). 

We focus on the U.S.\ state of Florida.
According to the 2020 decennial census, Florida has a population of approximately $21.5$ million, of which $20\%$ is below the age of 18 and $20\%$ is above the age of $65$. $50.8\%$ of the population is female, and $21\%$ of the population is foreign-born. $26.8\%$ of the population is Hispanic or Latino, $17\%$ is of African-American decent, and $3\%$ is Asian. $88.5\%$ of the population have at least a high-school diploma, and $30.5\%$ at least a Bachelor's degree. 
$12.4\%$ of the population is below the poverty line. 
Due to the diverse make-up of Florida and the fact that counties on the coast are expected to differ from neighboring inland counties in terms of income and other socio-demographic characteristics,
Florida serves as a useful running example.
Motivated by these considerations,
we demonstrate the proposed statistical framework by using 2019 ACS five-year period design-based estimates of log median household income for counties in Florida.

\section{Statistical framework} 
\label{section: Methodology}

In general,
we consider a spatial domain $\mD$,
a countable set:
e.g., 
the spatial domain $\mD$ may be the set of all $N = 67$ counties in Florida.
We are interested in data $Y_i$ indexed by spatial units $i \in \mD$:
e.g.,
$Y_i$ may be the log median household income for county $i \in \mD$ in Florida.
Many models for spatial data $Y_i$ assume that observations $Y_i$ and $Y_j$ are more similar when spatial units $i \in \mD$ and $j \in \mD$ are geographic neighbors,
e.g., 
when $i$ and $j$ are neighboring counties in Florida.
In other words,
if $B_{i,j} = 1$ indicates that spatial units $i$ and $j$ are geographic neighbors and $B_{i,j} = 0$ indicates that $i$ and $j$ are not neighbors,
then $Y_i$ and $Y_j$ are assumed to be more similar in the event $\{B_{i,j} = 1\}$ than in the event $\{B_{i,j} = 0\}$.

That said,
while geography may be important,
it may not be all-important:
e.g.,
counties on the coast of Florida are expected to differ from neighboring inland counties in terms of income,
because affluent residents may settle in counties on the coast while less affluent residents may be pushed into neighboring counties off the coast.
Therefore,
some geographical neighbors (e.g., neighboring counties on the coast) may be similar,
while other geographical neighbors (e.g., neighboring counties on the coast and off the coast) may be dissimilar.
As a consequence,
specifying neighborhood indicators $B_{i,j}$ based on geography alone can result in misspecified models.
Instead of specifying neighborhood indicators $B_{i,j}$,
we learn neighborhood indicators $B_{i,j}$ from spatial data $Y_i$ by adapting latent space models from the statistical analysis of {\em network data} to the statistical analysis of {\em spatial data.}

We divide the description of the proposed statistical framework into
\begin{itemize}
\item the {\em neighborhood model,}
that is,
the model that generates neighborhood indicators $B_{i,j}$ based on the positions of spatial units $i \in \mD$ and $j \in \mD$ in an unobserved socio-demographic space (Section~\ref{section: NNSD});\vspace{.2cm}
\item the {\em data model,} 
that is,
the model that generates spatial data $Y_i$ conditional on neighborhood indicators $B_{i,j}$ (Section~\ref{sec:data.model}).
\end{itemize}
To set the stage,
we first review latent space models for network data.

\subsection{Background: latent space models for network data}
\label{sec:background}

Latent space models for {\em network data} assume that the data consist of {\em observed} connection indicators $A_{i,j} \in \{0, 1\}$,
where the event $\{A_{i,j} = 1\}$ indicates that units $i$ and $j$ are connected:
e.g.,
the network may correspond to collaborations among researchers and $\{A_{i,j} = 1\}$ indicates that researchers $i$ and $j$ collaborate.
In one of the simplest scenarios,
connections are undirected and self-connections are excluded,
that is,
$A_{i,j} = A_{j,i}$ and $A_{i,i} = 0$ for all $i \neq j$.

\begin{figure}[t]
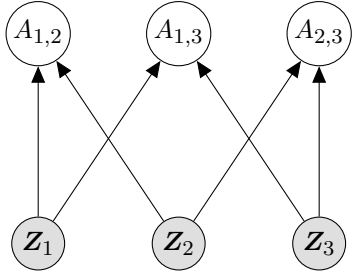

\center
      \tikz{ %
        \node[\obs] (y) {$A_{1,2}$} ; %
        \node[\obs, right=of y, yshift=0cm] (y2) {$A_{1,3}$} ; %
        \node[\obs, right=of y2, yshift=0cm] (y3) {$A_{2,3}$} ; %
        \node[\lat, below=of y, yshift=-1cm] (h1) {$\bm{Z}_1$} ; %
        \node[\lat, below=of y2, yshift=-1cm] (h3) {$\bm{Z}_2$} ; %
        \node[\lat, below=of y3, yshift=-1cm] (h4) {$\bm{Z}_3$} ; %
        \edge{h1} {y,y2} ; %
        \edge{h3} {y,y3} ; %
        \edge{h4} {y2,y3} ; %
      }
\caption{\label{fig:dag.lsm}
A directed acyclic graph representing the conditional independence structure of latent space models for network data ($A_{1,2}$, $A_{1,3}$, $A_{2,3}$). 
Unshaded circles indicate observable random variables ($A_{1,2}$, $A_{1,3}$, $A_{2,3}$), 
while shaded circles indicate unobservable random variables ($\bm{Z}_1$, $\bm{Z}_2$, $\bm{Z}_3$).
The directed acyclic graph reveals that the connection indicators $A_{1,2}$, $A_{1,3}$, $A_{2,3}$ are independent conditional on the positions $\bm{Z}_1$, $\bm{Z}_2$, $\bm{Z}_3$ of units $1, 2, 3$ in $\mR^q$.
}
\end{figure}

Euclidean latent space models \citep{Hoff2002} assume that units $i$ have positions $\bm{Z}_i$ in a $q$-dimensional Euclidean space $\mathbb{R}^q$ ($q \geq 1$).
In applications to social network data,
the latent space is sometimes interpreted as an unobserved social space.
Conditional on the positions $\bm{Z}_i = \bm{z}_i$ and $\bm{Z}_j = \bm{z}_j$ of units $i$ and $j$ in $\mathbb{R}^q$,
the connection indicators $A_{i,j} \in \{0, 1\}$ are independent Bernoulli random variables,
and the conditional probability of the event $\{A_{i,j} = 1\}$ depends on the Euclidean distance $|\!|\bm{z}_i-\bm{z}_j|\!|_2$ between units $i$ and $j$ in $\mathbb{R}^q$:
\beno    
\logit(\mbP(A_{i,j} = 1 \mid \alpha,\, \bm{Z}_i = \bm{z}_i,\, \bm{Z}_j = \bm{z}_j))     
&\coloneqq& \alpha - |\!|\bm{z}_i-\bm{z}_j|\!|_2.
\ee
The conditional independence structure of latent space models is represented in Figure~\ref{fig:dag.lsm}.
The intercept $\alpha\in \mathbb{R}$ controls the expected number of connections when the distances between all pairs of units $i$ and $j$ vanish.
A comprehensive review of latent space models can be found in \cite{Smith2019}.

We adapt latent space models from the statistical analysis of {\em network data} to the statistical analysis of {\em spatial data.}
It is worth noting that latent space models for {\em network data} assume that the network of interest is {\em observed,}
whereas the proposed latent space models for {\em spatial data} assume that the network of interest is {\em unobserved.}

\subsection{Neighborhood model} 
\label{section: NNSD}

\begin{figure}[t]
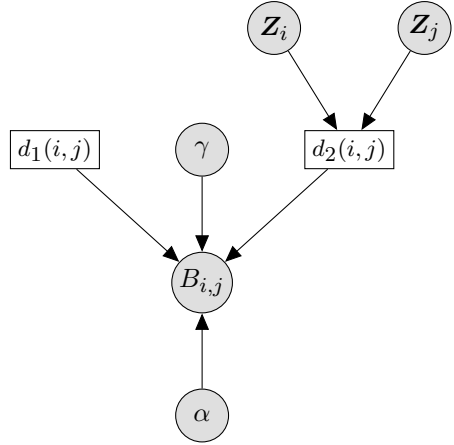

\center
  \tikz{ %
        \node[\lat] (y3) {$\gamma$};
        \node[rectangle, draw, left=of y3] (y) {$d_{1}(i,j)$} ; %
        \node[rectangle, draw, right=of y3, yshift=0cm] (y2) {$d_{2}(i,j)$} ; 
        \node[\lat, below=of y3, yshift=0cm] (P) {$B_{i,j}$} ; %
        \node[\lat, above=of y2, xshift=-1cm] (h1) {$\bm{Z}_i$} ; %
        \node[\lat, above=of y2, xshift=1cm] (h2) {$\bm{Z}_j$} ; %
        \node[\lat, below=of P, yshift=0cm, xshift=0cm] (g) {$\alpha$} ; 
        \edge{y2} {P} ; %
        \edge{y3} {P} ; %
        \edge{y} {P} ; %
        \edge{h1,h2} {y2} ;
        \edge{g} {P};
      } \label{fig: DAG}
\caption{The conditional independence structure of the neighborhood model that generates neighborhood indicators $B_{i,j}$.
Unshaded circles indicate observable random variables, 
while shaded circles indicate unobservable random variables.
Rectangles represent either known constants or known functions of random variables.}

      \label{fig: network}
\end{figure}

The neighborhood model generates neighborhood indicators $B_{i,j}$ conditional on the positions of spatial units $i \in \mD$ and $j \in \mD$ in an underlying socio-demographic space.
The conditional independence structure of the neighborhood model is represented by the directed acyclic graph in Figure~\ref{fig: network}.
The neighborhood model assumes that spatial units $i \in \mD$ have positions $\bm{Z}_i$ in an unobserved metric space,
which is taken to be $\mR^q$ and is interpreted as a socio-demographic space.
In the statistical analysis of {\em network data,}
it is known that selecting the dimension $q$ of $\mathbb{R}^q$ is possible but non-trivial \citep{latent.space.models.theory,LuChMc23},
and it is common practice to choose $q=2$ or $q = 3$ to facilitate visual representations.
In the statistical analysis of {\em spatial data,}
selecting the dimension $q$ of $\mR^p$ is more challenging,
because the network of interest is {\em unobserved.}
That being said,
it is natural to match the dimensions of the geographical and socio-demographic space:
e.g.,
in the application to the 2019 ACS data in Section~\ref{section: RealEg},
the geographical space is $\mR^2$ and it is hence natural to consider the socio-demographic space to be $\mR^2$.
We hence consider $q=2$.

The neighborhood model then assumes that,
conditional on the positions $\bm{Z}_i = \bz_i$ and $\bm{Z}_j = \bz_j$ of spatial units $i \in \mD$ and $j \in \mD$ in $\mR^2$,
neighborhood indicators $B_{i,j} \in \{0, 1\}$ are generated by drawing
\beno    
B_{i,j} \mid \alpha,\, \gamma,\, \mathbf{Z}_i = \bz_i,\, \mathbf{Z}_j = \bz_j 
&\ind& \hbox{Bernoulli}(p_{i,j}),
\ee
where the log odds of the conditional probability $p_{i,j}$ of the event $\{B_{i,j} = 1\}$ is
\begin{equation}
    \begin{array}{ccc}
\label{logit.a}
\logit(p_{i,j}) 
&\coloneqq& \alpha-\gamma\, d_1{(i, j)}-(1-\gamma)\, d_2(i, j).
\end{array}
\end{equation}
The intercept $\alpha\in \mathbb{R}$ controls the expected number of neighbors when the two terms $d_1(i,j)$ and $d_2(i,j)$ vanish,
while the weight $\gamma \in [0, 1]$ specifies the relative importance of $d_1(i, j)$ relative to $d_2(i, j)$.
The terms $d_1(i,j)$ and $d_2(i, j)$ represent the geographical and socio-demographic distance between spatial units $i \in \mD$ and $j \in \mD$,
respectively.
Both distances are assumed to be Euclidean:
e.g.,
in the application to the 2019 ACS data in Section~\ref{section: RealEg},
the geographical distance $d_1(i, j)$ is the Euclidean distance between the centroids of counties $i$ and $j$ in a two-dimensional geographical space and $d_2(i,j) \coloneqq |\!|\bz_i-\bz_j|\!|_2$ is the Euclidean distance between the positions $\bz_i$ and $\bz_j$ of counties $i$ and $j$ in a two-dimensional socio-demographic space.
The fact that the convex combination $\gamma\, d_1(i, j) + (1-\gamma)\, d_2(i,j)$ of the two distances $d_1(i, j)$ and $d_2(i,j)$ is subtracted from the intercept $\alpha$ implies that the probability of being neighbors decreases with increasing geographical distance $d_1(i, j)$ and increasing socio-demographic distance $d_2(i, j)$.
In other words,
the conditional probability of the event that spatial units $i \in \mD$ and $j \in \mD$ are neighbors is high when both $d_1(i, j)$ and $d_2(i, j)$ are small,
and is low when either distance is large.
Other specifications of the model are possible,
based on non-Euclidean distances and non-convex combinations of distances.
The proposed specification has the advantage of being simple and helping assess the relative importance of geographical distance $d_1(i, j)$ relative to socio-demographic distance $d_2(i, j)$.
It is worth noting that $d_1(i, j)$ is known while $d_2(i, j)$ is unknown, 
but $d_2(i, j)$ can be inferred from the posterior of positions $\bZ_1, \dots, \bZ_N$ given spatial data $Y_1, \dots, Y_N$.

It is convenient to use bivariate Gaussian priors for the positions $\bZ_i \in \mathbb{R}^2$:
\beno
\mathbf{Z}_i \mid \sigma^2_z\; \iid\; \hbox{MVN}_2(\bm{0}_2,\, \sigma^2_z\, \mathbf{I}_2),
\ee
where $\bm{0}_2 \in \mR^2$ denotes the two-dimensional null vector,
$\sigma^2_z > 0$ denotes a variance parameter,
and $\mathbf{I}_2 \in \{0, 1\}^{2 \times 2}$ denotes the $2 \times 2$-identity matrix.
If available,
covariates can be used as predictors of positions $\bm{Z}_i \in \mR^2$:
\begin{equation}\nonumber
    \mathbf{Z}_i \mid \mathbf{S}_i,\, \bm{\delta},\, \sigma^2_z\; \ind\; \hbox{MVN}_2 (\mathbf{S}_i\, \boldsymbol{\delta},\, \sigma^2_z\, \mathbf{I}_2),
\end{equation}
where $\mathbf{S}_i \in \mathbb{R}^{2 \times k}$ represents  covariates that help inform the position $\mathbf{Z}_i \in \mR^2$ of spatial unit $i \in \mD$ and $\boldsymbol{\delta}\in\mathbb{R}^k$ represents covariate weights ($k \geq 1$).

The neighborhood model (with covariates) can then be described as follows:
\beno
    B_{i,j} \mid \alpha,\, \gamma,\, \mathbf{Z}_i = \bz_i,\, \mathbf{Z}_j = \bz_j\; \ind\; \hbox{Bernoulli}(p_{i,j})\s 
\\
    \alpha \mid \sigma^2_{\alpha} \;\sim\; \hbox{Normal}(0,\, \sigma^2_{\alpha})\s
\\
    \gamma \;\sim\; \hbox{Uniform}(0,\, 1)\s
\\
    \mathbf{Z}_i \mid \mathbf{S}_i,\, \boldsymbol{\delta},\, \sigma^2_z \;\ind\; \hbox{MVN}_2 (\mathbf{S}_i\, \boldsymbol{\delta},\, \sigma^2_z\, \mathbf{I}_2)\s
\\
    \boldsymbol{\delta} \mid \sigma^2_{\delta} \;\sim\; \hbox{MVN}_k (\mathbf{0}_k,\, \sigma^2_{\delta}\, \mathbf{I}_k), 
\ee
where $p_{i,j}$ is specified by  \eqref{logit.a} and $\sigma^2_{\alpha} > 0$, $\sigma^2_{z} > 0$, and $\sigma^2_{\delta} > 0$ are hyperparameters that can be specified or can be assigned hyperpriors.

\paragraph*{Identifiability issues}

While $d_1(i, j)$ is observed and is thus fixed,
$d_2(i, j)$ is unobserved and is invariant to reflection, translation, and rotation of the positions $\bZ_i$ and $\bZ_j$ of spatial units $i \in \mD$ and $j \in \mD$ in $\mR^2$.
Such identifiability issues can be addressed by basing statistical inference on equivalence classes of positions \citep{Hoff2002}.
In addition,
to ensure that statistical conclusions regarding the weights $\gamma$ and $1-\gamma$ of the distances $d_1(i, j)$ and $d_2(i,j)$ are meaningful,
$d_1(i, j)$ and $d_2(i,j)$ need to be on the same scale.
We therefore restrict the positions of spatial units $i \in \mD$ to the closed unit disk in $\mathbb{R}^2$,
both in the geographical and socio-demographic space.

\subsection{Data model}
\label{sec:data.model}

\begin{figure}[t]
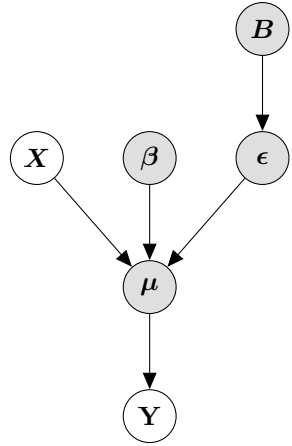

\centering
 \tikz{ %
        \node[\lat] (P) {$\bm{\mu}$} ; %
        \node[\lat, above=of P, yshift=0cm, xshift=0cm] (XX) {$\bm{\beta}$} ; %
        \node[\obs, below=of P, xshift=0cm] (Y) {$\mathbf{Y}$};
        \node[\lat, above=of P, yshift=0cm, xshift=1.5cm] (X) {$\bm{\epsilon}$} ; %
        \node[\obs, above=of P, xshift=-1.5cm] (e) {$\bm{X}$} ; %
        \node[\lat, above=of e, xshift=3cm] (D) {$\bm{B}$} ; %
        \edge{XX,X,e} {P} ; %
        \edge{D} {X} ;
        \edge{P} {Y};
      }
\caption{The conditional independence structure of the data model that generates spatial data $\bm{Y} \coloneqq (Y_i)$ conditional on neighborhood indicators $\bm{B} \coloneqq (B_{i,j})$;
note that the conditional independence structure of the neighborhood model that generates $\bm{B}$ is represented in Figure~\ref{fig: network}.
Unshaded circles indicate observable random variables, 
while shaded circles indicate unobservable random variables.
}
      \label{fig: MEM}
\end{figure}

The data model generates spatial data $Y_i$ conditional on neighborhood indicators $B_{i,j}$.
The conditional independence structure of the data model is described by the directed acyclic graph in Figure~\ref{fig: MEM}.
The data model assumes that spatial data $\bm{Y} \coloneqq (Y_i) \in \mR^N$ are generated conditional on neighborhood indicators $\bm{B} \coloneqq (B_{i,j}) \in \{0, 1\}^{\binom{N}{2}}$:
\begingroup
\allowdisplaybreaks
\begin{align*}
    & \mathbf{Y} \mid \bm{\mu},\, \sigma^2_y \;\sim\; \hbox{MVN}_N (\bm{\mu},\, \sigma^2_y\, \mathbf{I}_N)\s
\\
    & \bm{\mu} \mid \bm{X},\, \boldsymbol{\beta},\, \boldsymbol{\epsilon},\, \sigma_{\mu}^2 \;\sim\; \hbox{MVN}_N (\mathbf{X}\, \boldsymbol{\beta} + \boldsymbol{\epsilon},\, \sigma_{\mu}^2\,\mathbf{I}_N)\s
\\
    &\boldsymbol{\beta} \mid \sigma^2_{\beta} \;\sim\; \hbox{MVN}_p(\mathbf{0}_p,\, \sigma^2_{\beta}\, \mathbf{I}_p)\s
\\
    &\boldsymbol{\epsilon} \mid \bm{B},\, \sigma^2_{\epsilon} \;\sim\; \hbox{MVN}_N (\mathbf{0}_N,\, \sigma^2_{\epsilon}\, (\bm{D}(\bm{B})-\bm{B})^{-1})\s
    \\
    &\sigma_{\mu}^2 \mid a_{\mu},\, b_{\mu} \;\sim\; \hbox{Inverse-Gamma}(a_{\mu},\, b_{\mu})\s
\\
    &\sigma^2_{\epsilon} \mid a_{\epsilon},\, b_{\epsilon} \;\sim\; \hbox{Inverse-Gamma}(a_{\epsilon},\, b_{\epsilon}),
\label{eq. mixed}
\end{align*}
\endgroup
where $N$ denotes the number of spatial units $i \in \mD$,\,
$\bm{X} \in \mR^{N \times p}$ represents covariates, 
and $\bm{\beta} \in \mR^p$ represents covariate weights,
including an intercept $\beta_1$.
The matrix $\bm{D}(\bm{B}) \in \mR^{N \times N}$ is a diagonal matrix with the numbers of neighbors $D_i(\bm{B}) \coloneqq \sum_{j = 1, j \neq i}^N\, B_{i,j}$ of spatial units $i \in \mD$ on its main diagonal.
The parameterization of the precision matrix $\boldsymbol{\Sigma}_{\epsilon}^{-1} \coloneqq (\bm{D}(\bm{B})-\bm{B})\, /\, \sigma_{\epsilon}^2$ of random effects vector $\boldsymbol{\epsilon} \in \mR^N$ resembles the parameterization of precision matrices in the popular ICAR models \citep{BesagYorkMollie1991}.  
While ICAR models specify neighborhood indicators $\bm{B}$ based on geography,
we learn $\bm{B}$ from spatial data $\bm{Y}$.
The parameters $\sigma^2_y > 0$,
$\sigma^2_{\beta} > 0$,
$a_{\mu} > 0$,
$b_{\mu} > 0$,
$a_{\epsilon} > 0$,
and $b_{\epsilon} > 0$ 
are hyperparameters that can be specified or can be assigned hyperpriors.
In the application to the 2019 ACS data in Section~\ref{section: RealEg},
the variance $\sigma^2_y > 0$ is based on the sampling error variance reported by the U.S.\ Census Bureau and the delta method.

\paragraph*{Identifiability issues}

The data model has identifiability issues,
in that the intercept $\beta_1$ and the random effect vector $\bm{\epsilon}$ in $\mathbf{X}\, \boldsymbol{\beta} + \boldsymbol{\epsilon}$ cannot be estimated unless additional constraints are imposed.
We impose the constraint $\sum_{i=1}^N \epsilon_i=0$ \citep[e.g.,][]{Porter2014b,Keefe2018}.

\subsection{Implementation}

We pursue a Bayesian approach,
choosing priors to ensure conjugacy of the neighborhood and data model.
We set 
$\sigma^2_{\alpha}=3$,
$\sigma^2_{\beta}=100$, 
$\sigma^2_{\delta}=100$, 
$\sigma^2_z=1$,  
$a_{\mu} = a_{\epsilon} = 2$, 
and $b_{\mu} = b_{\epsilon} = 1$,
so that the priors of the variance parameters are vague.
We use the Stan programming language \citep{Carpenter2017} to generate 20,000 draws from the posterior of interest,
discarding the first 10,000 draws as burn-in.

\subsection{Convergence diagnostics}

In addition to trace plots,
we use the multivariate Gelman-Rubin potential scale reduction factor (PSRF) of \citet{vats:18a} as a convergence diagnostic.
The multivariate Gelman-Rubin PSRF can be viewed as a stable version of the Gelman-Rubin convergence diagnostic \citep{GaRd92} and has the advantage of providing a principled approach to detecting non-convergence.

\section{Simulation studies}
\label{section: Simulations}

We conduct two simulation studies.
The first one is based on simulating data that resemble the 2019 ACS data.
The second one is based on simulating data from the proposed statistical framework.
Throughout Sections \ref{section: Simulations} and \ref{section: RealEg},
we refer to the proposed statistical framework as the Nearest-Neighbor Socio-Demographic (NNSD) network model. 
The special case of the NNSD network model with $\gamma = 0$ is called the SD network model, 
and the special case with $\gamma=1$ is referred to as the NN model.
We compare the proposed NNSD, SD, and NN network models with the traditional ICAR models,
which assume that two spatial units are neighbors if the units are geographical neighbors (e.g., neighboring counties). 

\subsection{Empirical simulation study} 
\label{section: EmpSimulations}

We first conduct an empirical simulation study based on simulated data that resemble the 2019 ACS data used in Section~\ref{section: RealEg}.
We generate 50 pseudo-datasets by sampling
\beno
\mathbf{Y}^\star \mid \mathbf{Y} = \mathbf{y},\, \boldsymbol{\Sigma}_y &\sim& \hbox{MVN}_N(\mathbf{y},\, \boldsymbol{\Sigma}_y),
\ee
where $\mathbf{y}$ represents the 2019 ACS data described in Section~\ref{section: Data}.
The resulting pseudo-data are centered at the 2019 ACS data,
with variance-covariance matrix $\mathbf{\Sigma}_y$ equal to the 2019 ACS sampling error variance-covariance matrix. 
We use the natural logarithm of the 2019 ACS design-based estimates of median housing cost as a covariate. 
In addition,
we include an intercept term in the model by adding a column of $1$'s to the matrix $\mathbf{X}$.
We do not use covariates to predict the latent positions $\mathbf{Z}_i\in \mathbb{R}^2$ of spatial units $i \in \mD$ in the socio-demographic space,
and use the prior
\beno
\mathbf{Z}_i \mid \sigma_z^2 = 1\; \iid\; \hbox{MVN}_2(\bm{0}_2,\, \sigma_z^2\, \mathbf{I}_2).
\ee
We compare posterior estimates (i.e., posterior means) in terms of mean-squared error (MSE) and mean absolute error (MAE),
defined as 
\beno
MSE 
\coloneqq \displaystyle\frac{1}{N}\sum\limits_{i=1}^N (Y^\star_i-\widehat{Y}^\star_i)^2\ \ \mbox{ and } \ \ \ \hbox{MAE} \coloneqq \displaystyle\frac{1}{N} \sum\limits_{i=1}^N |Y^\star_i-\widehat{Y}^\star_i|,
\ee
where $\widehat{Y}^\star_i$ denotes a posterior median of spatial unit $i \in \mD$. 

\begin{table}[b]
\caption{Empirical simulation study based on data $\mathbf{Y}^\star$ that are generated conditional on the 2019  ACS data $\mathbf{Y} = \mathbf{y}$.}
\setlength{\tabcolsep}{10pt}
\centering
\begin{tabular}{c|l|l}
\multicolumn{1}{ c| }{Network model} & \multicolumn{1}{ c| }{MSE} & \multicolumn{1}{ c }{MAE}\\
 \cline{1-3}\noalign{\vskip 1mm}   
\multicolumn{1}{ l| }{ICAR} & $7.18\times 10^{-4}$ & $6.55\times 10^{-2}$ \\
\multicolumn{1}{ l| }{NN} & $7.43\times 10^{-5}$ & $1.58\times 10^{-2}$ \\
\multicolumn{1}{ l| }{NNSD} & $4.60\times 10^{-5}$ & $1.44\times 10^{-2}$\\ 
\multicolumn{1}{ l| }{SD}  & $6.45\times 10^{-5}$ & $1.57\times 10^{-2}$ \\ 

\cline{1-3}
 \end{tabular}\s

\label{table: ESS}
\end{table}
Table~\ref{table: ESS} shows the median MSE and MAE across the $50$ pseudo-datasets and suggests that the NNSD network model produces the closest estimates to the pseudo data with the lowest MAE and MSE among all models.
The NNSD network model outperforms the ICAR model by a large margin, 
and the NN and SD network models by a small margin.
That makes sense,
because the NNSD network model is the most general model and contains the NN and SD network models as special cases.

Figure~S1 in the supplement shows the design-based estimates of median household income for counties in Florida and an instance of a pseudo-dataset,
on the log scale. 
Figure~S1 suggests that the pseudo-dataset resembles the 2019 ACS data. 
In addition,
we present posterior estimates using the NNSD,
NN,
and SD network models as well as the ICAR model.  
We see that the proposed NNSD network models improve estimates compared with the traditional ICAR network model, 
with the NNSD network model estimates being closest to the pseudo-dataset. 
The posterior estimates using all three network models are able to capture the underlying spatial trend in the pseudo-dataset.
The proposed NNSD network models produces comparable estimates to the traditional ICAR model, 
but with much lower MSE and MAE,
as is evident from Table~\ref{table: ESS}.
As a consequence,
incorporating socio-demographic information helps reduce uncertainty in model-based predictions,
compared with traditional nearest-neighbor models (e.g., the traditional ICAR model). 

\subsection{Model-based simulation study}

The model-based simulation study complements the empirical simulation study in Section \ref{section: EmpSimulations} by simulating data from the proposed NNSD network model and its special cases,
the NN and SD network models.
In contrast to the pseudo-data generated in the empirical simulation study,
the data generated in the model-based simulation study need not resemble the 2019 ACS data used in Section~\ref{section: RealEg}.

We consider three scenarios:
the NNSD network model with $\gamma = 1$ (Scenario 1),
which is equivalent to the NN network model;
the NNSD network model with $\gamma = 0.5$ (Scenario 2);
and the NNSD network model with $\gamma = 0$ (Scenario 3),
which is equivalent to the SD network model. 
To generate data,
we use the geographical distances $d_1(i,j)$ between the centroids of counties $i \in \mD$ and $j \in \mD$ in Florida, 
while the socio-demographic distances $d_2(i, j)$ between $i$ and $j$ are generated by sampling positions $\mathbf{Z}_i$ and $\mathbf{Z}_j$ using {\tt R} package {\tt latentnet} \citep{latentnet.jss}.
We then generate neighborhood indicators $B_{i,j}$ by sampling
\beno
B_{i,j} \mid \alpha,\, \gamma,\, \bm{Z}_i = \bz_i,\, \bm{Z}_j = \bz_j &\ind& \hbox{Bernoulli}(p_{i,j}),
\ee
where $p_{i,j}$ is given by \eqref{logit.a} and the weights $\alpha$ and $\gamma$ are given by $\alpha=-2.5$ and $\gamma \in \{0,\, 0.5,\, 1\}$.
Conditional on $\bm{B}$,
we generate 50 datasets by sampling
\beno
\boldsymbol{\epsilon} \mid \bm{B},\, \sigma^2_{\epsilon}=1\; \sim\; \hbox{MVN}_N(\mathbf{0}_N,\, \sigma^2_{\epsilon}\, (\bm{D}(\bm{B})-\bm{B})^{-1})\s
\\
\bm{\mu} \mid \bm{X},\, \bm{\beta} = (-6.20,\, 2.5)^\top,\; \bm{\epsilon},\; \sigma_{\mu}^2=0.12 \;\sim\; \hbox{MVN}_N (\mathbf{X}\, \boldsymbol{\beta} + \boldsymbol{\epsilon},\, \sigma_{\mu}^2\, \mathbf{I}_N)\s
\\
\mathbf{Y} \mid \bm{\mu},\, \sigma^2_y=0.12 \;\sim\; \hbox{MVN}_N (\bm{\mu},\, \sigma^2_y\, \mathbf{I}_N),
\ee
where $\bm{D}(\bm{B}) \in \mR^{N \times N}$ is a diagonal matrix with the numbers of neighbors $D_i(\bm{B}) \coloneqq \sum_{j = 1, j \neq i}^N\, B_{i,j}$ of spatial units $i \in \mD$ on its main diagonal.
The matrix of covariates $\mathbf{X} \in \mR^{2 \times N}$ consists of a column of $1$'s and a column of the 2019 ACS design-based estimates of median housing cost,
on the log scale.
As posterior estimates,
we use posterior means based on 10,000 post-burn-in samples from Stan. 

\begin{table}[t]
\caption{\label{table: Sims2}
Model-based simulation results based on data generated under Scenarios~1, 2, and 3.}
\begin{subtable}[!htb]{1.0\textwidth}
\caption{Scenario 1: NNSD with $\gamma = 1$, equivalent to NN}
\setlength{\tabcolsep}{10pt}
\centering
\begin{tabular}{c|l|l}
\multicolumn{1}{ c| }{Network model}  & \multicolumn{1}{ c| }{MSE} & \multicolumn{1}{ c }{MAE}\\ 
\cline{1-3} \noalign{\vskip 1mm}  
\multicolumn{1}{ l| }{ICAR} & $6.76\times 10^{-1}$ & $6.44\times10^{-1}$ \\ 
\multicolumn{1}{ l| }{NN} & $3.34\times 10^{-5}$ & $1.33\times10^{-2}$ \\ 
\multicolumn{1}{ l| }{NNSD} & $2.27\times 10^{-5}$ & $1.19\times 10^{-2}$\\ 
\multicolumn{1}{ l| }{SD} & $1.36\times 10^{-4}$ & $2.02\times 10^{-2}$ \\ 
 \end{tabular}
\end{subtable}

\begin{subtable}[!htb]{1.0\textwidth}
\setlength{\tabcolsep}{10pt}
\caption{Scenario 2: NNSD with $\gamma = 0.5$}
\centering
\begin{tabular}{c|l|l}
\multicolumn{1}{ c| }{Network model}  & \multicolumn{1}{ c| }{MSE} & \multicolumn{1}{ c }{MAE}\\ 
\cline{1-3}\noalign{\vskip 1mm}   
\multicolumn{1}{ l| }{ICAR} & $6.64\times 10^{-1}$ & $6.52\times 10^{-1}$ \\ 
\multicolumn{1}{ l| }{NN} & $2.39\times 10^{-5}$ & $1.24\times10^{-2}$\\ 
\multicolumn{1}{ l| }{NNSD} & $2.19\times 10^{-5}$ & $1.18\times 10^{-2}$ \\ 
\multicolumn{1}{ l| }{SD} & $2.33\times 10^{-5}$ & $1.20\times 10^{-2}$\\ 
 \end{tabular}
 \label{table: Simps2.b}
\end{subtable}

\begin{subtable}[!htb]{1\textwidth}
\caption{Scenario 3: NNSD with $\gamma = 0$, equivalent to SD}
\setlength{\tabcolsep}{10pt}
\centering
\begin{tabular}{c|l|l}
\multicolumn{1}{ c| }{Network model}  & \multicolumn{1}{ c| }{MSE} & \multicolumn{1}{ c }{MAE}\\ 
\cline{1-3}\noalign{\vskip 1mm}   
\multicolumn{1}{ l| }{ICAR} & $7.06\times 10^{-1}$ & $6.58\times 10^{-1}$ \\ 
\multicolumn{1}{ l| }{NN} & $4.74\times 10^{-4}$ & $2.34\times 10^{-2}$ \\ 
\multicolumn{1}{ l| }{NNSD} & $3.46\times 10^{-5}$  & $1.42\times 10^{-2}$ \\ 
\multicolumn{1}{ l| }{SD} & $1.11\times 10^{-4}$  & $1.95\times 10^{-2}$ \\ 
 \end{tabular}
 \label{table: Simps2.c}
\end{subtable}

\end{table}

Table~\ref{table: Sims2} shows the results of the model-based simulation studies. 
In all scenarios, 
the proposed NNSD network model and its special cases (the NN and SD network models) outperform the classic ICAR model in terms of MSE and MAE. 
In Scenario~1, 
where the NN network model generates data based on geography,
the NNSD network model produces the model-based estimates with the lowest MSE and MAE, 
followed by the NN network model. 
Both the NNSD and NN network models outperform the SD network model,
which ignores geography.
In Scenario 2, 
where the NNSD network model generates data based on both geographical and socio-demographic distance,
all three network model estimates are comparable, with the NNSD network estimates performing slightly better than the NN and SD network estimates.   
In Scenario 3, 
where the SD network model generates data,
posterior analysis suggests that the NNSD network outperforms the SD network. 
This observation dovetails with the observation made in Section~\ref{section: EmpSimulations} and makes sense, 
because the NNSD network model contains the NN and SD network models as special cases. 

An instance of generated data is shown in Figure~S2 in the supplement,
and is compared with the model estimates under the ICAR, NNSD, NN, and SD network models.
We see that the estimates based on the ICAR model are smoother compared with the NNSD, NN, and SD network models and fail to capture the spatial trends in the generated data,
whereas the estimates based on the NNSD, NN, and SD network models are close to the generated data and capture the counties and regions with high median household income. 

The results of the model-based simulation study reinforce the findings from the empirical simulation study, 
wherein we conclude that the proposed NNSD network model improves the model estimates both in terms of MSE and MAE. 
Moreover,
while the traditional ICAR model produces smoother estimates, 
the proposed NNSD network produces more precise estimates as it learns the spatial trends from the geographical location of the counties as well as the latent positions of the counties in the underlying socio-demographic space. 

\section{Application to 2019 ACS data} 
\label{section: RealEg}

We use the NNSD network model to predict the median household income of counties in Florida on the log scale,
using the 2019 ACS design-based estimates of the median household income of counties in Florida on the log scale as response variable $\mathbf{Y}$.
To help estimate median household income, 
we use the 2019 ACS design-based estimates of the median housing cost for counties in Florida on the log scale as a covariate $\mathbf{X}$.
In addition, 
we include an intercept term by adding a column of $1$'s in $\mathbf{X}$. 
The latent positions $\bZ_i$ of units $i$ in the socio-demographic space are estimated using the design-based estimates of the percentage of county population below the poverty line as a covariate $\bm{S}_i$. 
We assume that the sampling error variance $\sigma_y^2$ of the response variable $\mathbf{Y}$ is known, 
based on the sampling error variance reported by the U.S.\ Census Bureau in combination with the delta method. We generate 20,000 draws from the posterior using Stan \citep{Carpenter2017}, discarding the first 10,000 samples as burn-in.  
Posterior estimates (i.e., posterior medians) are based on the post-burn-in draws from the posterior. 
We use the NNSD network model along with the SD and NN network models to generate model-based estimates. 
As a benchmark, 
we use the traditional ICAR model.

\begin{sidewaysfigure}[!htpb]
    \centering
    \includegraphics[width=\columnwidth]{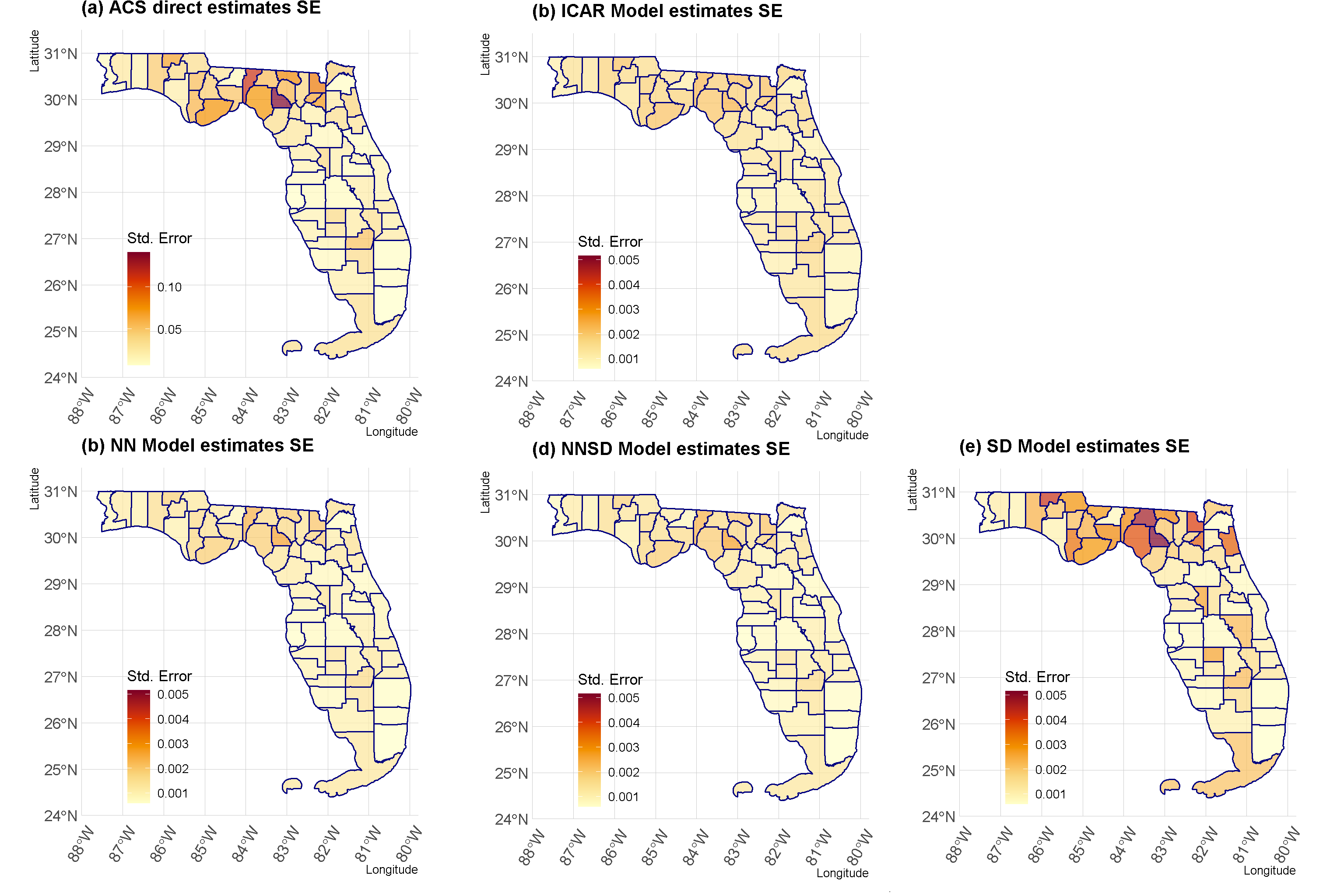}
\caption{2019 ACS data application: maps of Florida showing the standard errors (SE) of 
(a) the 2019 ACS design-based estimates,
(b) the ICAR estimates,
(c) the NN network model estimates,
(d) the NNSD network model estimates,
and (e) the SD network model estimates.
The SE of the design- and model-based estimates are presented on different scales.
}
    \label{fig: RD_se}
\end{sidewaysfigure}

Figure~S3 in the supplement compares the model-based estimates obtained under the three network models against the design-based estimates of median household income for counties in Florida,
on the log scale. 
We see that the model-based estimates resemble the ACS design-based estimates. 
The estimates based on the NNSD, NN, and SD network models are able to capture the spatial trends evident in the design-based estimates
and emulate the spatial trends captured by the more traditional ICAR model. 
The posterior estimate of $\gamma$ is $0.51$, 
giving substantial weight to both geographical and socio-demographic distances,
although there appears to be considerable uncertainty about $\gamma$ (95\% posterior credible interval $[0.01,\, 0.98]$).

\begin{figure}[t]
    \centering
    \includegraphics[width=\columnwidth,height=0.95\linewidth]{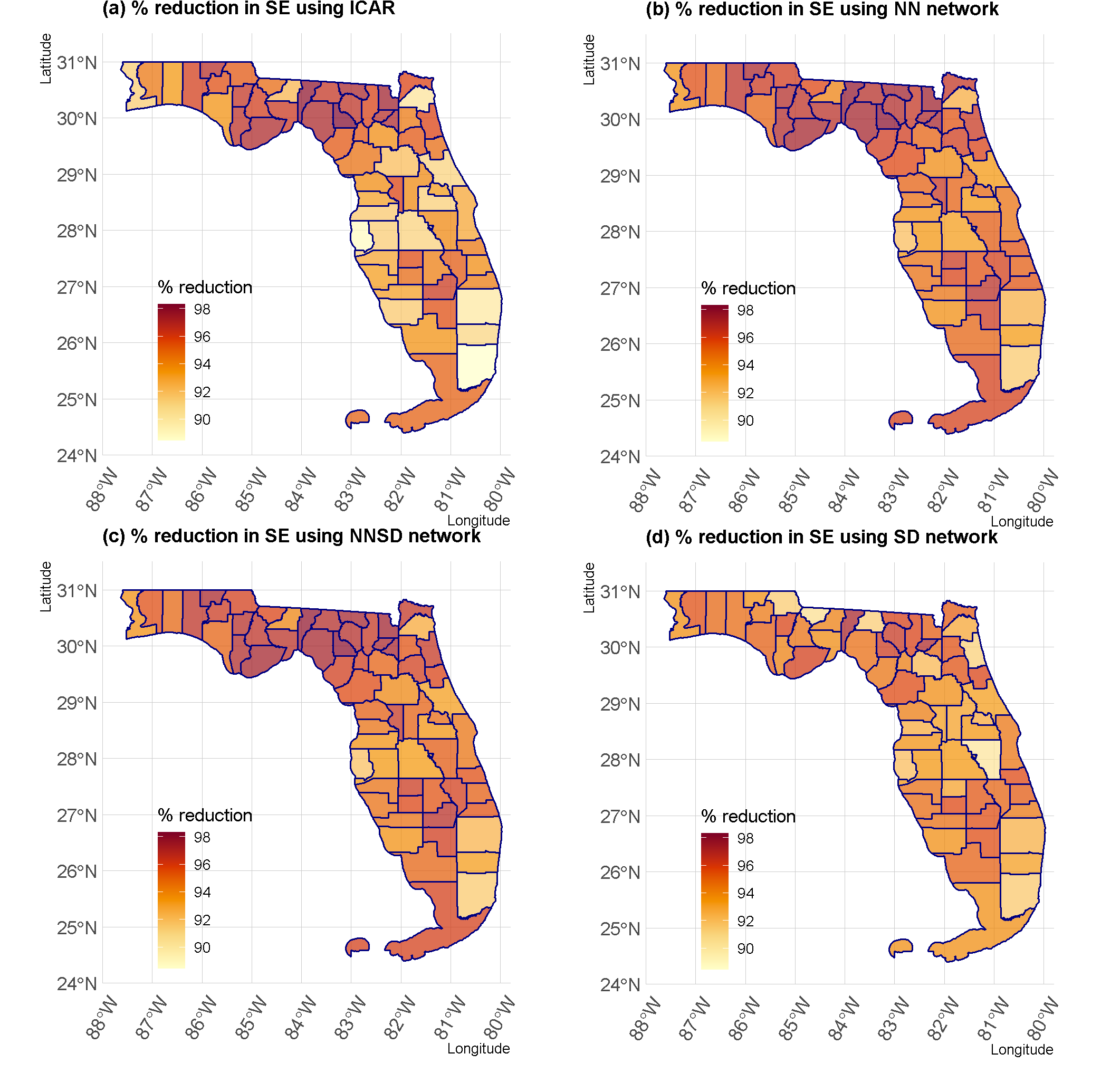}
\caption{2019 ACS data application: maps of Florida showing the percentage
reduction in standard errors achieved by 
(a) the ICAR model,
(b) the NN network model,
(c) the NNSD network model,
and (d) the SD network model,
over the standard errors of the 2019 ACS design-based estimates.}
    \label{fig: RD_se_perc_reduction}
\end{figure}

Figure~\ref{fig: RD_se} presents standard errors of the design-based estimates and compares them with the posterior standard errors of the model-based estimates under the three network models as well as the ICAR model. 
We observe that all models provide a substantial reduction in standard errors compared with the design-based estimates. 
\hide{
Michael: The following is repetitive. We disucss that in the following paragraphs in more detail. I therefore outcommented the following.

Among the three network models, 
the SD network model gives rise to the highest standard errors,
compared with the NNSD and NN network models.
We observe that even in the counties in northern Florida and the Florida Panhandle where the design-based estimates have high standard errors, 
both the NNSD and NN network models reduce posterior uncertainty.
}

In Figure~\ref{fig: RD_se_perc_reduction} we plot the percentage reduction in posterior standard errors based on the NNSD, SD, and NN network models over the standard errors of the design-based estimates,
with reductions ranging from $90\%$ to $98\%$.
The NNSD and NN network models show comparable percentage reductions, 
both outperforming the SD network model and the ICAR model. 
It is worth noting that the ICAR model reports the lowest reduction in areas with the most populous and diverse cities,
including Miami, Fort Lauderdale, Orlando, Tampa, and Jacksonville. 

\begin{figure}[t]
    \centering
\includegraphics[width=\columnwidth]{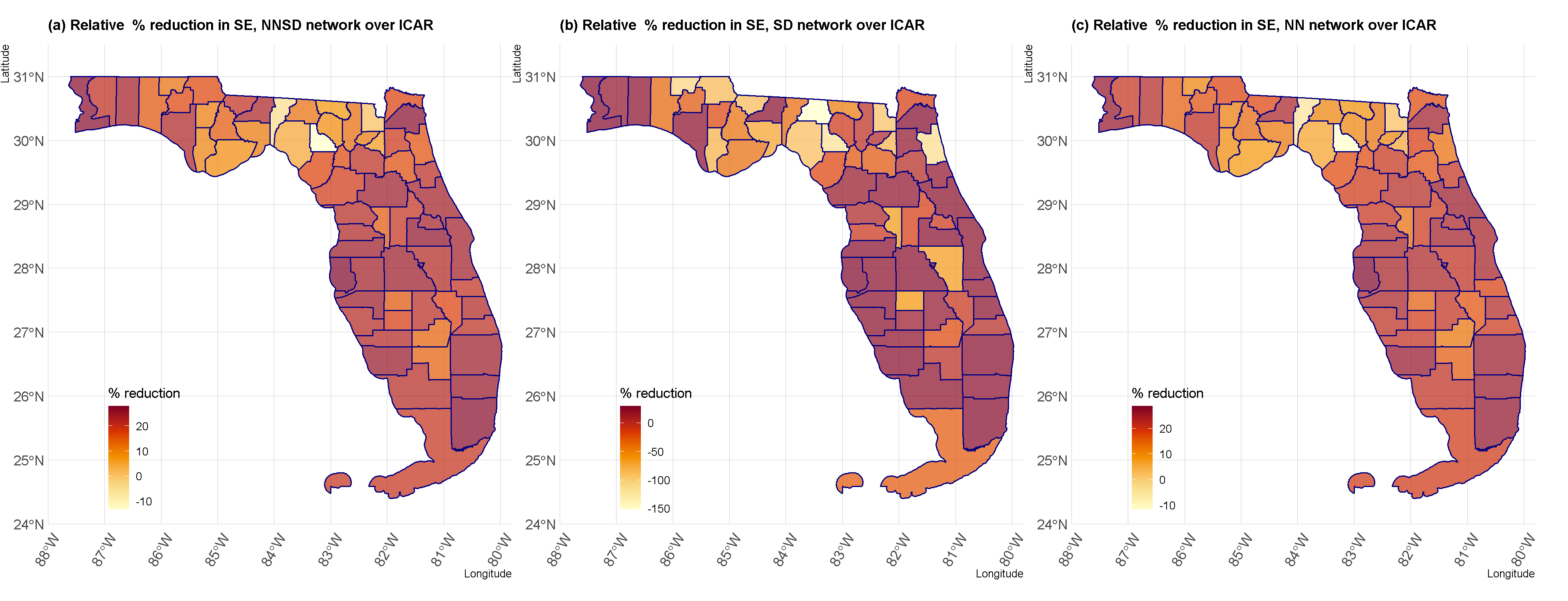}
\caption{2019 ACS data application: maps of Florida showing the relative percentage reduction in standard errors achieved by (a) the NNSD network model,
(b) the SD network model, 
and (c) the NN network model, 
over the ICAR model. 
Observe that different plots have different color scales.}

    \label{fig: RD_se_rel_perc_reduction}
\end{figure}

We map the relative percentage reduction in standard errors of the three network models,
relative to the ICAR model,
in Figure~\ref{fig: RD_se_rel_perc_reduction}. 
The NNSD network model achieves an average percentage reduction of 15-20\% in standard errors over the ICAR model.
By contrast,
the SD network model performs worse than the ICAR model,
which suggests that ignoring geography and estimating neighborhoods based on socio-demographic differences alone may be insufficient for emulating key features of spatially correlated data.
Taken together,
these two observations demonstrate that,
when the neighborhood structure is not known with certainty or cannot be specified based on geography alone,
network models that account for both geographical and socio-demographic distances can be beneficial. 

\begin{figure}[t]
    \centering
\includegraphics[width=0.8\columnwidth]{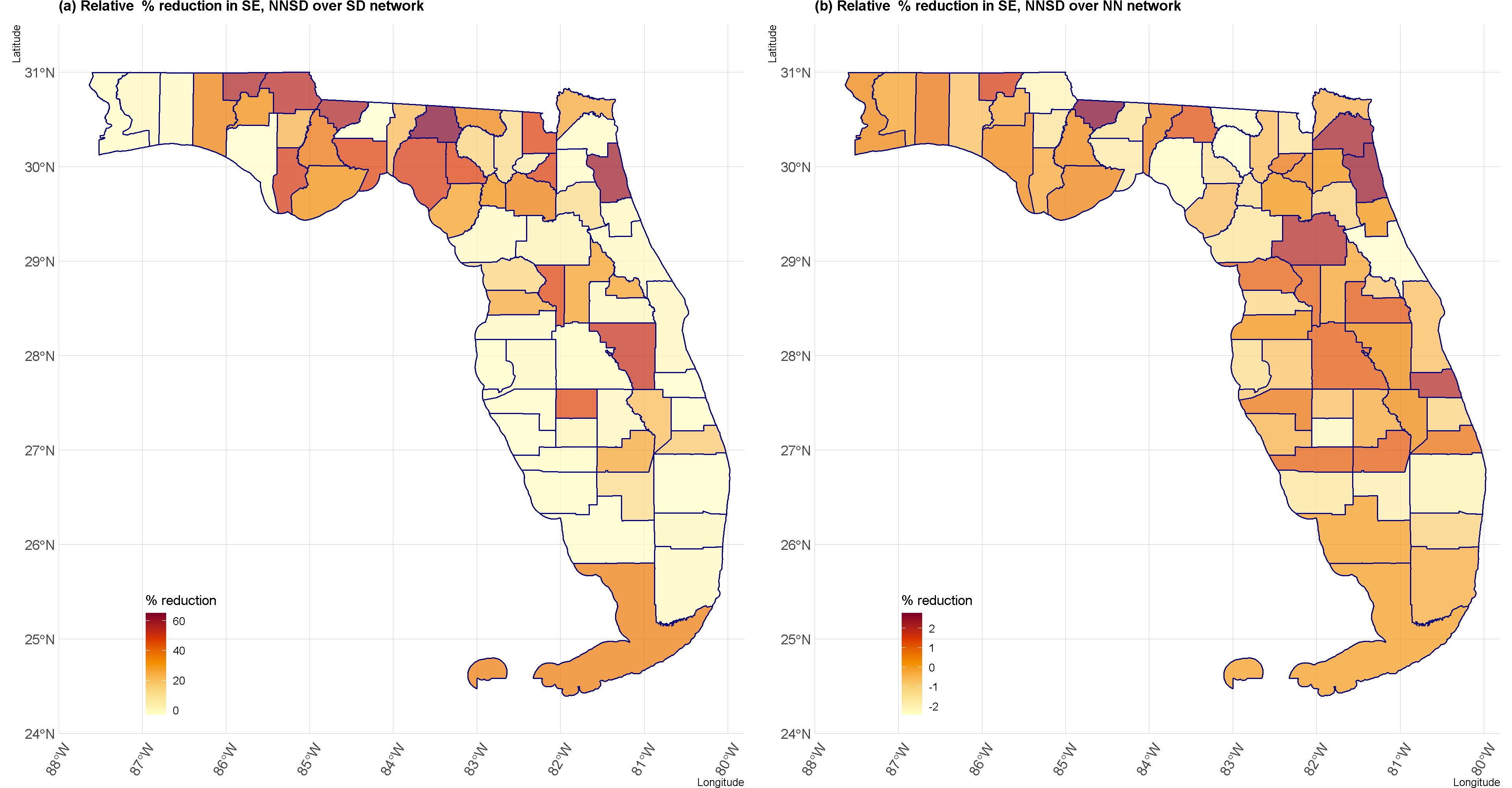}
\caption{2019 ACS data application: maps of Florida showing the relative percentage reduction in standard errors achieved by 
(a) the NNSD network model, 
over the SD network model,
and (b) the NNSD network model, 
over the NN network model.
Observe that different plots have different color scales.}
    \label{fig: RD_se_rel_perc_reduction2}
\end{figure}

In Figure~\ref{fig: RD_se_rel_perc_reduction2} we map the relative percentage reduction in standard errors among the three network models. 
The average relative percentage reduction in the standard errors of the NNSD network against the SD network is about $40\%$.
It is interesting to note that the highest reductions in standard errors occur in northern Florida and the Florida Panhandle,
where geography may be more important than in the coastal areas of southern Florida with socio-demographic differences between counties on the coast and neighboring counties off the coast.
The average relative percentage reduction in standard errors of the NNSD network model compared with the NN network model is about $2\%$.
These observations underscore that geography is important but not all-important,
and that models incorporating both geographical and socio-demographic distances help improve model estimates while reducing posterior uncertainty.
The added value of socio-demographic space,
which is visible even in the limited geographical area of Florida,
is expected to be more pronounced in larger and more heterogenous spatial domains (e.g., entire continents).

\section{Discussion} 
\label{section: Discussion}

We have introduced a novel approach to learning neighborhoods from spatial data,
which is useful when geography is important but not all-important owing to geographical,
political,
and social divisions.
The proposed approach may prove useful in scenarios where survey data fall short.
For example,
traditional survey data---including the 2019 ACS data---are typically reported at higher levels of geography, 
where the number of observations used to construct the estimates produce reliable results.
However,
these same data may fail to produce reliable estimates at lower levels of geography.
In such scenarios,
the proposed approach may help produce more reliable small area estimates than conventional approaches,
because the socio-demographic space can capture features at lower levels of geography.
A possible application is the study of the spread of infectious diseases through communities.

Many open problems remain.
For example,
we mentioned in Section~\ref{sec:introduction} that (1) geographical non-neighbors can be more similar than would be expected based on geographical distance; 
and (2) geographical neighbors can be dissimilar due to geographical,
political,
and social divisions.
We focused on issue (2) while leaving issue (1) to future research.
An interesting direction for future research is to tackle both issues (1) and (2).

\begin{acks}[Acknowledgments]
This research was partially supported by the U.S.~National Science Foundation under NSF awards SES-1853096, NCSE-2215168, and DMS-1812119. This article is released to inform interested parties of ongoing research and to encourage discussion. The views expressed on statistical issues are those of the authors and not those of the NSF or U.S. Census Bureau.
\end{acks}

\begin{supplement}
The supplement includes Figures~S1,
S2,
and S3 referenced in Sections~\ref{section: Simulations} and \ref{section: RealEg}.
\end{supplement}

\bibliographystyle{asa1}

\bibliography{reference}

\begin{thebibliography}{33}
\newcommand{\enquote}[1]{``#1''}
\expandafter\ifx\csname natexlab\endcsname\relax\def\natexlab#1{#1}\fi

\bibitem[{Athreya et~al.(2021)Athreya, Tang, Park, and
  Priebe}]{latent.space.models.theory}
Athreya, A., Tang, M., Park, Y.,  and Priebe, C.~E. (2021), \enquote{On
  estimation and inference in latent structure random graphs,}
  \textit{Statistical Science}, 36, 68--88.

\bibitem[{Bartlett(1967)}]{Bartlett1967}
Bartlett, M.~S. (1967), \enquote{Inference and stochastic processes,}
  \textit{Journal of the Royal Statistical Society. Series A (General)},
  457--478.

\bibitem[{Besag(1974)}]{Besag1974}
Besag, J. (1974), \enquote{Spatial interaction and the statistical analysis of
  lattice systems,} \textit{Journal of the Royal Statistical Society: Series B
  (Methodological)}, 36, 192--225.

\bibitem[{Besag et~al.(1991)Besag, York, and Mollié}]{BesagYorkMollie1991}
Besag, J., York, J.,  and Mollié, A. (1991), \enquote{Bayesian image
  restoration, with two applications in spatial statistics,} \textit{Annals of
  the Institute of Statistical Mathematics}, 43, 1--20.

\bibitem[{Carpenter et~al.(2017)Carpenter, Gelman, Hoffman, Lee, Goodrich,
  Betancourt, Brubaker, Guo, Li, and Riddell}]{Carpenter2017}
Carpenter, B., Gelman, A., Hoffman, M.~D., Lee, D., Goodrich, B., Betancourt,
  M., Brubaker, M., Guo, J., Li, P.,  and Riddell, A. (2017), \enquote{Stan: A
  probabilistic programming language,} \textit{Journal of Statistical
  Software}, 76, 1--32.

\bibitem[{Chaskin(1997)}]{Chaskin1997}
Chaskin, R.~J. (1997), \enquote{Perspectives on neighborhood and community: A
  review of the literature,} \textit{Social Service Review}, 71, 521--547.

\bibitem[{Chetty et~al.(2018)Chetty, Friedman, Hendren, Jones, and
  Porter}]{Chetty2018}
Chetty, R., Friedman, J.~N., Hendren, N., Jones, M.~R.,  and Porter, S.~R.
  (2018), \enquote{The opportunity atlas: Mapping the childhood roots of social
  mobility,} Tech. rep., National Bureau of Economic Research.

\bibitem[{Christiansen et~al.(2022)Christiansen, Baumann, Kuemmerle, Mahecha,
  and Peters}]{ChBaKuMaPe22}
Christiansen, R., Baumann, M., Kuemmerle, T., Mahecha, M.~D.,  and Peters, J.
  (2022), \enquote{Toward causal inference for spatio-temporal data: {C}onflict
  and forest loss in {C}olombia,} \textit{Journal of the American Statistical
  Association}, 117, 591--601.

\bibitem[{Csisz\'ar and Talata(2006)}]{Csiszar2006}
Csisz\'ar, I.,  and Talata, Z. (2006), \enquote{Consistent estimation of the
  basic neighborhood of {M}arkov random fields,} \textit{The Annals of
  Statistics}, 34, 123--145.

\bibitem[{Cutchin et~al.(2011)Cutchin, Eschbach, Mair, Ju, and
  Goodwin}]{Cutchin2011}
Cutchin, M.~P., Eschbach, K., Mair, C.~A., Ju, H.,  and Goodwin, J.~S. (2011),
  \enquote{The socio-spatial neighborhood estimation method: an approach to
  operationalizing the neighborhood concept,} \textit{Health \& Place}, 17,
  1113--1121.

\bibitem[{Galster(2001)}]{Galster2001}
Galster, G. (2001), \enquote{On the nature of neighbourhood,} \textit{Urban
  Studies}, 38, 2111--2124.

\bibitem[{Gao and Bradley(2019)}]{GaO2019}
Gao, H.,  and Bradley, J.~R. (2019), \enquote{Bayesian analysis of areal data
  with unknown adjacencies using the stochastic edge mixed effects model,}
  \textit{Spatial Statistics}, 31, 100357.

\bibitem[{{Gelman} and {Rubin}(1992)}]{GaRd92}
{Gelman}, A.,  and {Rubin}, D.~B. (1992), \enquote{Inference from iterative
  simulation using multiple sequences,} \textit{Statistical Science}, 7,
  457--472.

\bibitem[{Grannis(1998)}]{Grannis1998}
Grannis, R. (1998), \enquote{The importance of trivial streets: Residential
  streets and residential segregation,} \textit{American Journal of Sociology},
  103, 1530--1564.

\bibitem[{Grannis(2005)}]{Grannis2005}
--- (2005), \enquote{T-communities: Pedestrian street networks and residential
  segregation in Chicago, Los Angeles, and New York,} \textit{City \&
  Community}, 4, 295--321.

\bibitem[{{Hoff} et~al.(2002){Hoff}, {Raftery}, and {Handcock}}]{Hoff2002}
{Hoff}, P.~D., {Raftery}, A.~E.,  and {Handcock}, M.~S. (2002), \enquote{Latent
  space approaches to social network analysis,} \textit{Journal of the American
  Statistical Association}, 97, 1090--1098.

\bibitem[{Ji and Seymour(1996)}]{JiSe96}
Ji, C.,  and Seymour, L. (1996), \enquote{A consistent model selection
  procedure for {M}arkov random fields based on penalized pseudolikelihood,}
  \textit{The Annals of Applied Probability}, 6, 423--443.

\bibitem[{Keefe et~al.(2018)Keefe, Ferreira, and Franck}]{Keefe2018}
Keefe, M., Ferreira, M.,  and Franck, C. (2018), \enquote{On the formal
  specification of sum-zero constrained intrinsic conditional autoregressive
  models,} \textit{Spatial Statistics}, 24.

\bibitem[{Krivitsky and Handcock(2008)}]{latentnet.jss}
Krivitsky, P.~N.,  and Handcock, M.~S. (2008), \enquote{{Fitting position
  latent cluster models for social networks with {latentnet}},} \textit{Journal
  of Statistical Software}, 24, 1--23.

\bibitem[{Lu et~al.(2007)Lu, Reilly, Banerjee, and Carlin}]{Lu2007}
Lu, H., Reilly, C.~S., Banerjee, S.,  and Carlin, B.~P. (2007),
  \enquote{Bayesian areal wombling via adjacency modeling,}
  \textit{Environmental and ecological statistics}, 14, 433--452.

\bibitem[{Lubold et~al.(2023)Lubold, Chandrasekhar, and McCormick}]{LuChMc23}
Lubold, S., Chandrasekhar, A.~G.,  and McCormick, T.~H. (2023),
  \enquote{Identifying the latent space geometry of network models through
  analysis of curvature,} \textit{Journal of the Royal Statistical Society:
  Series B (Statistical Methodology)}, 1--63, to appear.

\bibitem[{Ma et~al.(2010)Ma, Carlin, and Banerjee}]{Ma2010}
Ma, H., Carlin, B.~P.,  and Banerjee, S. (2010), \enquote{Hierarchical and
  joint site-edge methods for Medicare hospice service region boundary
  analysis,} \textit{Biometrics}, 66, 355--364.

\bibitem[{Meinshausen and B\"uhlmann(2006)}]{MeBu06}
Meinshausen, N.,  and B\"uhlmann, P. (2006), \enquote{High-dimensional graphs
  and variable selection with the {LASSO},} \textit{The Annals of Statistics},
  34, 1436--1462.

\bibitem[{Porter et~al.(2014)Porter, Holan, Wikle, and Cressie}]{Porter2014b}
Porter, A.~T., Holan, S.~H., Wikle, C.~K.,  and Cressie, N. (2014),
  \enquote{Spatial {F}ay–{H}erriot models for small area estimation with
  functional covariates,} \textit{Spatial Statistics}, 10, 27--42.

\bibitem[{Ravikumar et~al.(2010)Ravikumar, Wainwright, and
  Lafferty}]{Ravikumar2010}
Ravikumar, P., Wainwright, M.~J.,  and Lafferty, J. (2010),
  \enquote{High-dimensional {I}sing model selection using $\ell_1$-regularized
  logistic regression,} \textit{The Annals of Statistics}, 38, 1287--1319.

\bibitem[{Schweinberger et~al.(2017)Schweinberger, Babkin, and
  Ensor}]{Schweinberger2017}
Schweinberger, M., Babkin, S.,  and Ensor, K.~B. (2017),
  \enquote{High-dimensional multivariate time series with additional
  structure,} \textit{Journal of Computational and Graphical Statistics}, 26,
  610--622.

\bibitem[{Smith et~al.(2019)Smith, Asta, and Calder}]{Smith2019}
Smith, A.~L., Asta, D.~M.,  and Calder, C.~A. (2019), \enquote{The geometry of
  continuous latent space models for network data,} \textit{Statistical
  Science}, 34, 428--453.

\bibitem[{Tobler(1970)}]{Tobler1970}
Tobler, W.~R. (1970), \enquote{A computer movie simulating urban growth in the
  {D}etroit region,} \textit{Economic Geography}, 46, 234--240.

\bibitem[{USCB(2020)}]{Census2020}
USCB (2020), \enquote{Understanding and using American Community Survey data:
  What all data users need to know,} Tech. rep., U.S.\ Department of Commerce,
  Economics and Statistics Administration.

\bibitem[{Vats and Knudson(2021)}]{vats:18a}
Vats, D.,  and Knudson, C. (2021), \enquote{Revisiting the Gelman-Rubin
  Diagnostic,} \textit{Statistical Science}, 36, 518–529.

\bibitem[{White and Ghosh(2009)}]{White2009}
White, G.,  and Ghosh, S.~K. (2009), \enquote{A stochastic neighborhood
  conditional autoregressive model for spatial data,} \textit{Computational
  Statistics \& Data Analysis}, 53, 3033--3046.

\bibitem[{Whittle(1954)}]{Whittle1954}
Whittle, P. (1954), \enquote{On stationary processes in the plane,}
  \textit{Biometrika}, 434--449.

\bibitem[{Wikle et~al.(2019)Wikle, Zammit-Mangion, and Cressie}]{WiZMCr19}
Wikle, C.~K., Zammit-Mangion, A.,  and Cressie, N. (2019),
  \textit{Spatio-Temporal Statistics with R}, Boca Raton, FL: Chapman \&
  Hall/CRC.

\end{thebibliography}

\newpage


\setcounter{page}{1}
\begin{appendix}

\begin{center}
\LARGE
{Supplementary Material:\\ 
A Socio-Demographic Latent Space Approach to Spatial Data When Geography is Important but Not All-Important}
\end{center}

\setcounter{figure}{0}
\renewcommand{\thefigure}{S\arabic{figure}}
\setcounter{table}{0}
\renewcommand{\thetable}{S\arabic{table}}

\begin{sidewaysfigure}[!htpb]
\begin{adjustwidth}{0cm}{}
    \centering
    \includegraphics[width=\columnwidth]{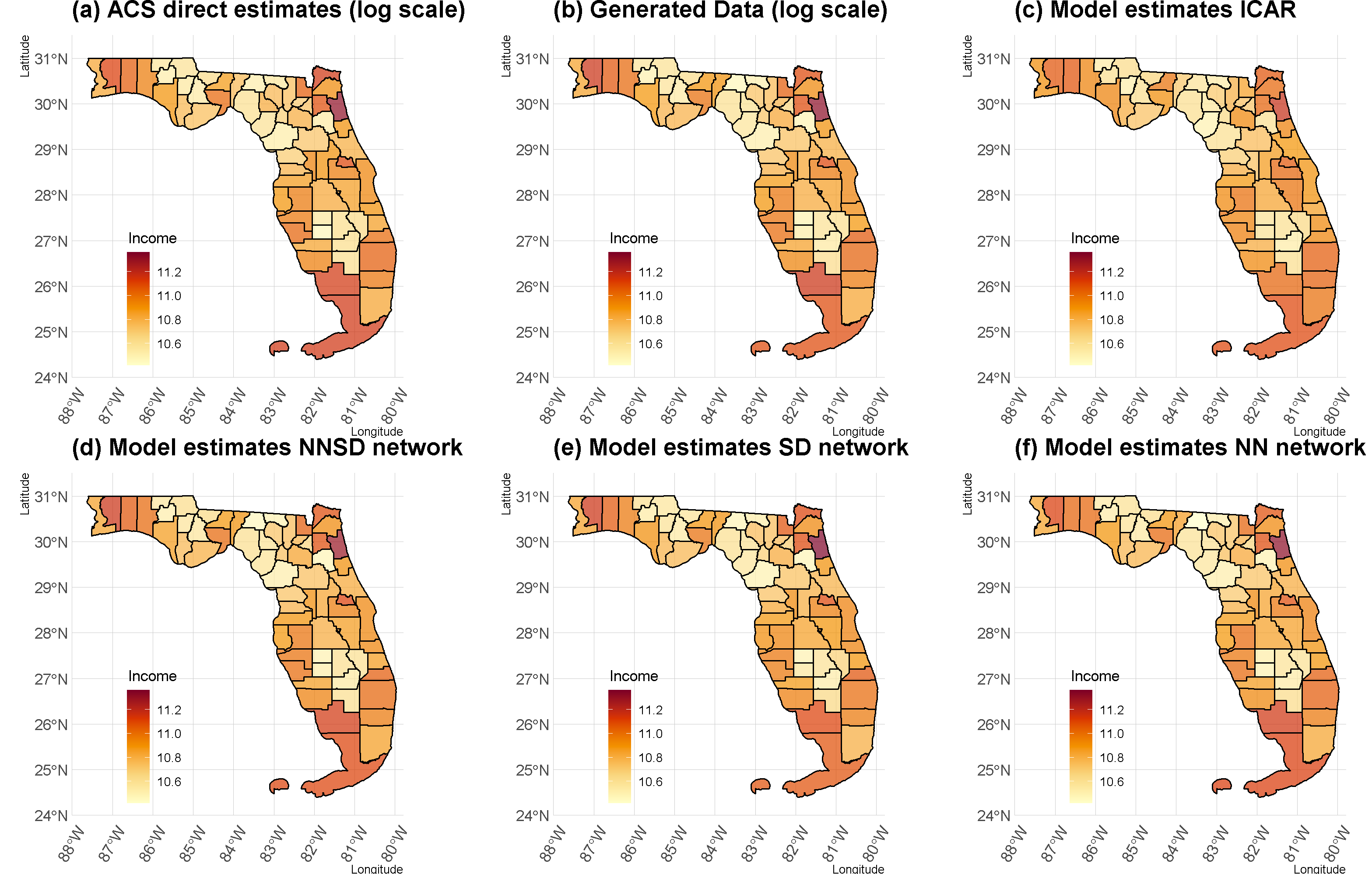}
\end{adjustwidth}
\caption{Empirical simulation results based on 
(a) the 2019 ACS design-based estimates $\mathbf{y}$ of the median household income of counties in Florida, 
on the log scale; 
(b) an instance of a pseudo-dataset $\mathbf{Y}^\star$ generated conditional on $\mathbf{Y} = \mathbf{y}$; 
(c) posterior estimates $\widehat{\mathbf{Y}}^\star$ using the ICAR model;
(c) posterior estimates $\widehat{\mathbf{Y}}^\star$ using the NNSD network model; 
(d) posterior estimates $\widehat{\mathbf{Y}}^\star$ using the SD network model;
and (e) posterior estimates $\widehat{\mathbf{Y}}^\star$ using the NN network model.}
\label{fig: ESS}
\end{sidewaysfigure}

\begin{sidewaysfigure}[!htpb]
\centering
\begin{subfigure}{\paperwidth}
    \centering
    \caption{\label{fig: SimpStudy2.a}}
    \includegraphics[width=\columnwidth]{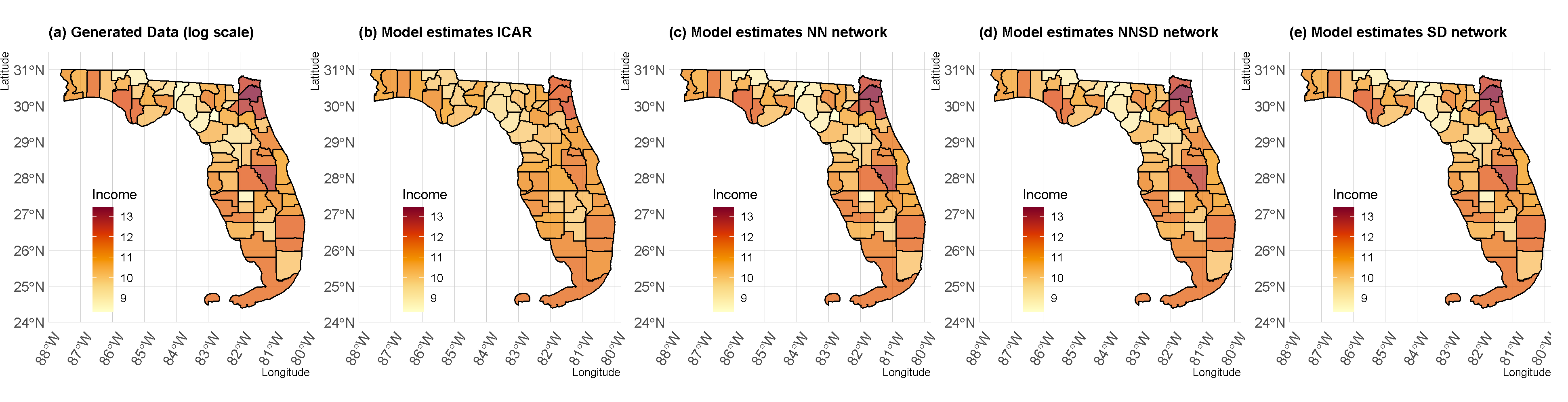}
\end{subfigure}\\
\begin{subfigure}{\paperwidth}
\centering
\caption{\label{fig: SimpStudy2.b}}
    \includegraphics[width=\columnwidth]{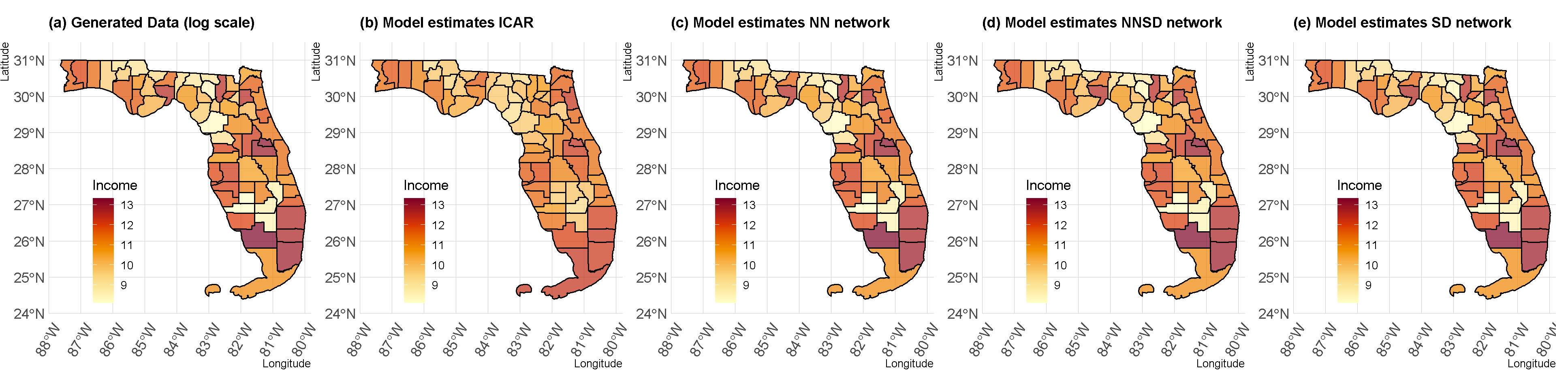}
\end{subfigure}\\
\begin{subfigure}{\paperwidth}
\centering
\caption{\label{fig: SimpStudy2.c}}
    \includegraphics[width=\columnwidth]{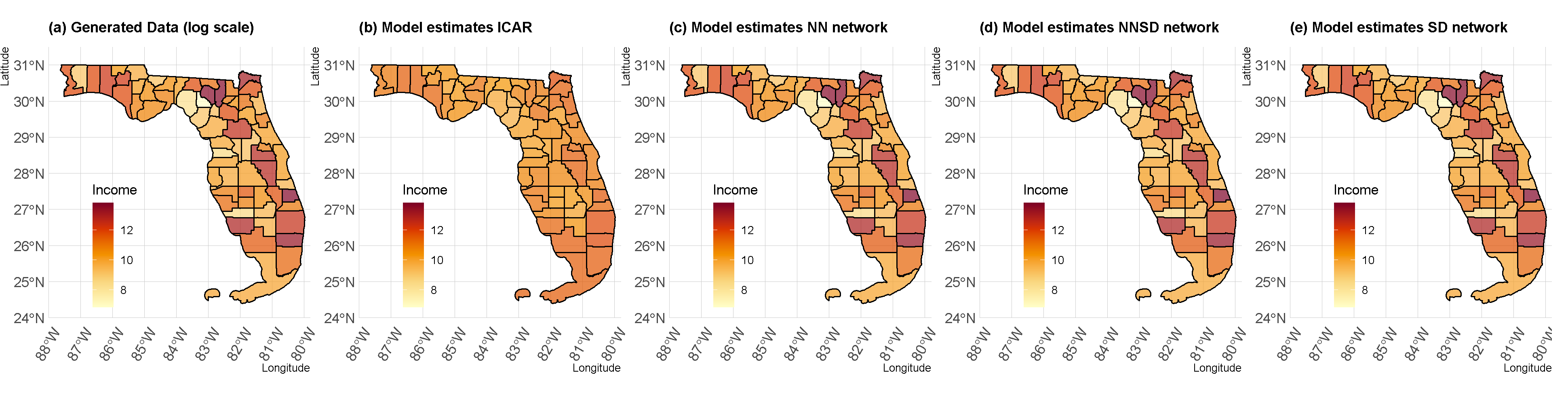}
\end{subfigure}%
\caption{Model-based simulation results:
maps of Florida comparing an instance of the generated data under 
(a) Scenario 1 $(\gamma=1)$,
(b) Scenario 2 $(\gamma=0.5)$,
and (c) Scenario 3 $(\gamma=0)$ with estimates based on the ICAR, NNSD, SD, and NN network models.}
\label{fig: Simp2}
\end{sidewaysfigure}

\begin{sidewaysfigure}[!htpb]
    \centering
    \includegraphics[width=\columnwidth]{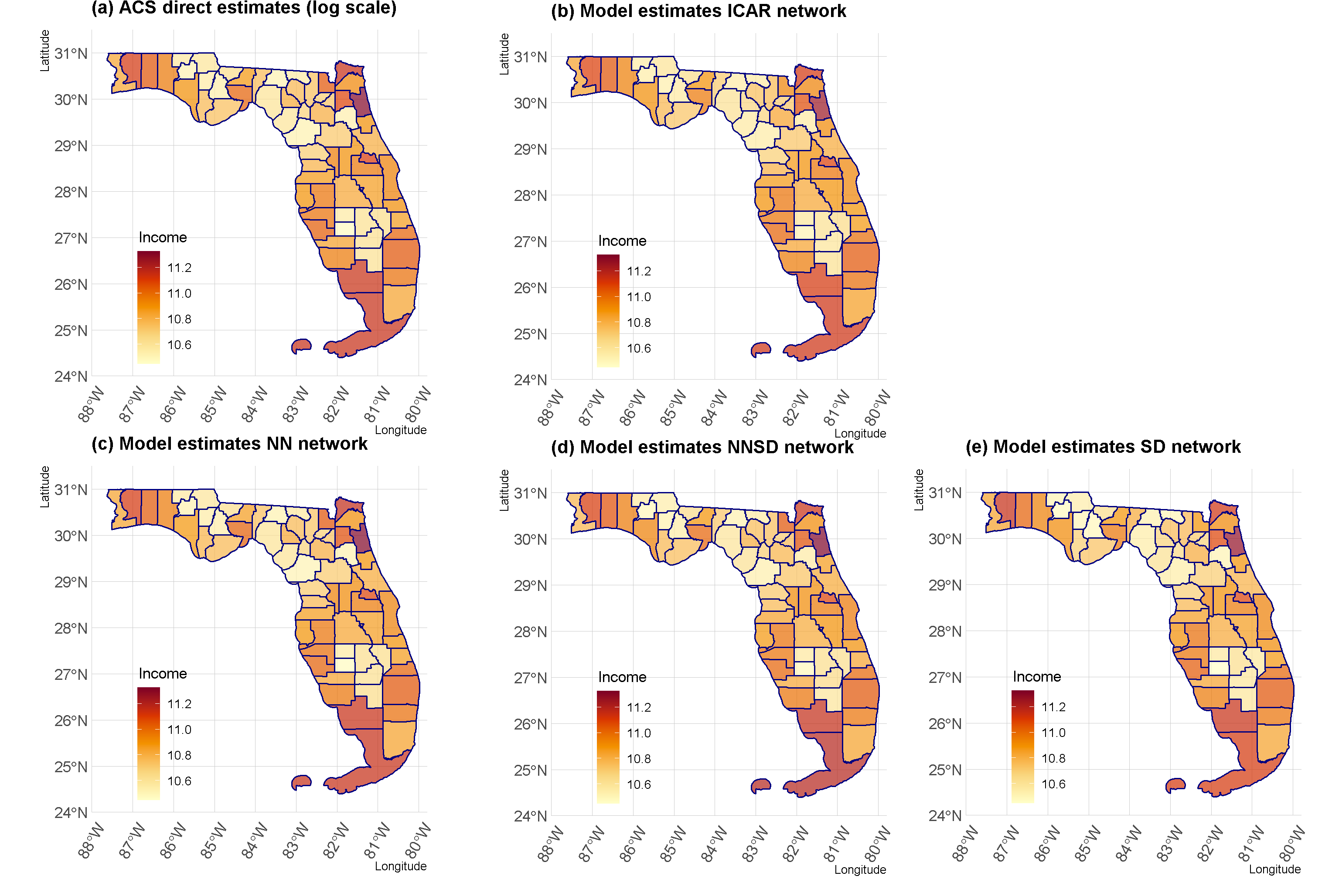}
\caption{2019 ACS data application: maps of Florida showing 
(a) the 2019 ACS design-based estimates of median household income per county, on the log scale; 
(b) the ICAR model estimates;
(c) the NN model estimates;
(d) the NNSD model estimates;
and (e) the SD model estimates.}
    \label{fig: RD_est}
\end{sidewaysfigure}

\end{appendix}

\end{document}